\documentclass[aps,prl,twocolumn,showpacs,superscriptaddress,amsmath,amssymb,longbibliography]{revtex4-1}
\usepackage{graphicx}
\usepackage{color}
\usepackage{float}
\usepackage[usenames,dvipsnames,svgnames,table]{xcolor}
\usepackage{dcolumn}% Align table columns on decimal point
\usepackage{bm}%bold math
\usepackage{algorithm}
\usepackage{algorithmicx}
\usepackage{algpseudocode}%
\usepackage{makecell}
\usepackage{array}
\usepackage{soul}
\usepackage{amsmath}
\usepackage{placeins}

\usepackage{url}

\usepackage{pdfpages}
\makeatletter
\AtBeginDocument{\let\LS@rot\@undefined}
\makeatother

\begin{document}

\title{Finding the optimal nets for self-folding Kirigami}

  \author{N. A. M. Ara\'ujo}
   \email{nmaraujo@fc.ul.pt}
    \affiliation{Departamento de F\'{\i}sica, Faculdade de Ci\^{e}ncias, Universidade de Lisboa, 
    1749-016 Lisboa, Portugal}
    \affiliation{Centro de F\'{i}sica Te\'{o}rica e Computacional, Universidade de Lisboa, 
    1749-016 Lisboa, Portugal}

 \author{R. A. da Costa}
 \affiliation{Department of Physics \& I3N, University of Aveiro, 3810-193 Aveiro, Portugal}
 
 \author{S. N. Dorogovtsev}
 \affiliation{Department of Physics \& I3N, University of Aveiro, 3810-193 Aveiro, Portugal}
 \affiliation{A. F. Ioffe Physico-Technical Institute, 194021 St. Petersburg, Russia}
 
 \author{J. F. F. Mendes}
 \affiliation{Department of Physics \& I3N, University of Aveiro, 3810-193 Aveiro, Portugal}

\pacs{05.65.+b,64.60.aq}

\begin{abstract}
Three-dimensional shells can be synthesized from the spontaneous self-folding
	of two-dimensional templates of interconnected panels, called nets.
	However, some nets are more likely to self-fold into the
	desired shell under random movements. The optimal nets are the ones
	that maximize the number of vertex connections, i.e., vertices that
	have only two of its faces cut away from each other in the net.
	Previous methods for finding such nets are based on random search and
	thus do not guarantee the optimal solution. Here, we propose a
	deterministic procedure. We map the connectivity of the shell into a
	shell graph, where the nodes and links of the graph represent the
	vertices and edges of the shell, respectively. Identifying the nets
	that maximize the number of vertex connections corresponds to finding
	the set of maximum leaf spanning trees of the shell graph. This method
	allows not only to design the self-assembly of much larger shell
	structures but also to apply additional design criteria, as a complete
	catalog of the maximum leaf spanning trees is obtained.
\end{abstract}

\maketitle

The synthesis of three-dimensional polyhedral shells at the micron and nano
scales is key for encapsulation and drug
delivery~\cite{Fernandes2012,Shim2012,Filippousi2013}. Inspired by the Japanese
art of Kirigami, where hollowed structures are obtained from cutting and
folding a sheet of paper, lithographic methods have been developed to form
shells from two-dimensional templates of interconnected
panels~\cite{Sussman2015,Zhang2015,Lamoureux2015,Shyu2015,Collins2016,Jacobs2016}.
The potential is enormous, for a wide range of shapes and sizes can be
obtained. Ideally, the unfolded templates (nets) should spontaneously self-fold
into the target structure to reduce production costs and achieve large-scale
parallel production.

\begin{figure}[t]
\centering
\includegraphics[width=0.9\columnwidth]{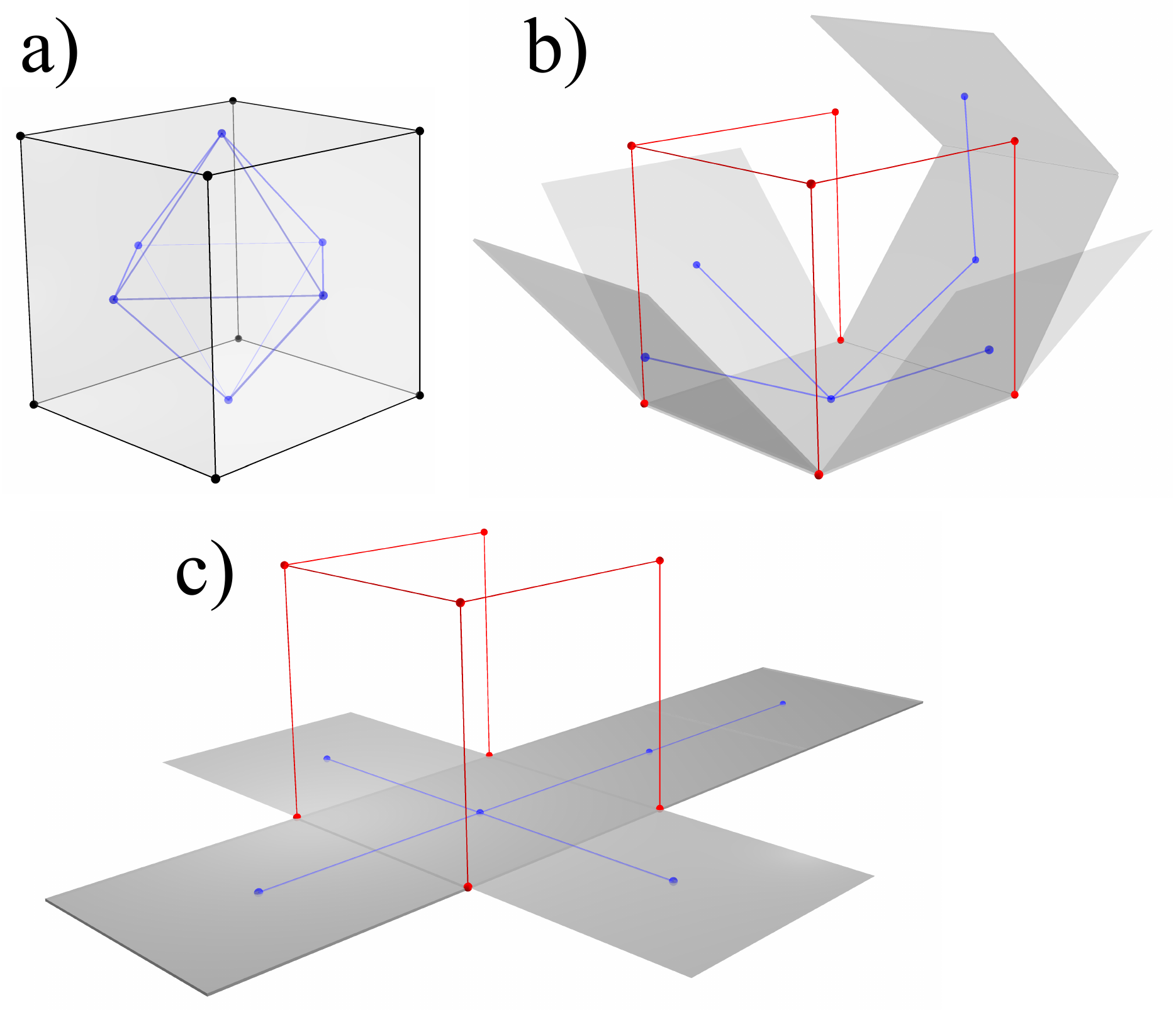}
	\caption{Net of a cubic shell. (a) The cubic shell is mapped into a
	\textit{shell graph} (black), where nodes and links of the
	\textit{shell graph} are the vertices and edges of the polyhedron,
	respectively. In the \textit{face graph} (blue), the nodes are the
	shell faces and the links connect pairs of adjacent faces. To unfold
	the shell into a two-dimensional template (net), one needs to remove a
	set of shell edges (e.g., red links in (b) and (c)).  This set of
	removed shell edges is a sub-graph of the shell graph. The set of
	removed shell edges (cut) and the net are spanning trees of the shell
	and face graphs, respectively. The four vertices of the bottom face are
	vertex connections.\label{fig::net_cube}}
\end{figure}
Many nets fold into the same structure, but some are more likely to self-fold
under random movements than others. Finding the optimal net for self-folding is
not trivial, as the self-folding pathway depends on the geometry of the net,
its physical properties, and its interactions with the surrounding
medium~\cite{Whitesides2002,Leong2007,Azam2011,azam2009compactness}.
Pandey~\textit{et al.} combined experimental and numerical techniques to
identify design principles for self-folding~\cite{Pandey2011}. They have
considered experimentally several nets as templates consisting of metallic
hinges and panels on the submillimeter scale and let them self-fold driven by
surface tension.  The self-folding efficiency of each net was quantified by
their yield, corresponding to the fraction of samples that self-folded into the
target polyhedron, free of defects.  They found that the yield is maximized
when, from the entire set of nets, one picks the nets with the maximum number
of so-called vertex connections (topological compactness)~\cite{Pandey2011}. A
vertex connection is a vertex shared in the polyhedron by two adjacent faces
that share, in the net, this vertex but not an edge, see
Fig.~\ref{fig::net_cube}. Searching for this global optimum in
Ref.~\cite{Pandey2011} implies considering all possible nets of the shell.
This inefficient procedure is time consuming, demanding, and for most shapes
impossible, since, as we show here, the total number of nets rapidly grows with
the number of edges.  For example, before symmetry reductions, with only twelve
edges, a cube has $384$ possible nets, while a dodecahedron, with thirty edges,
has more than $5$ million possible nets.  Consequently, methods to identify
optimal nets of sufficiently large shells have been based on random searches in
the configuration space.  They consider only a portion (actually a small
portion) of possible nets and so they do not guarantee optimal solutions.

\begin{figure*}[ht]
\centering
\includegraphics[width=\textwidth]{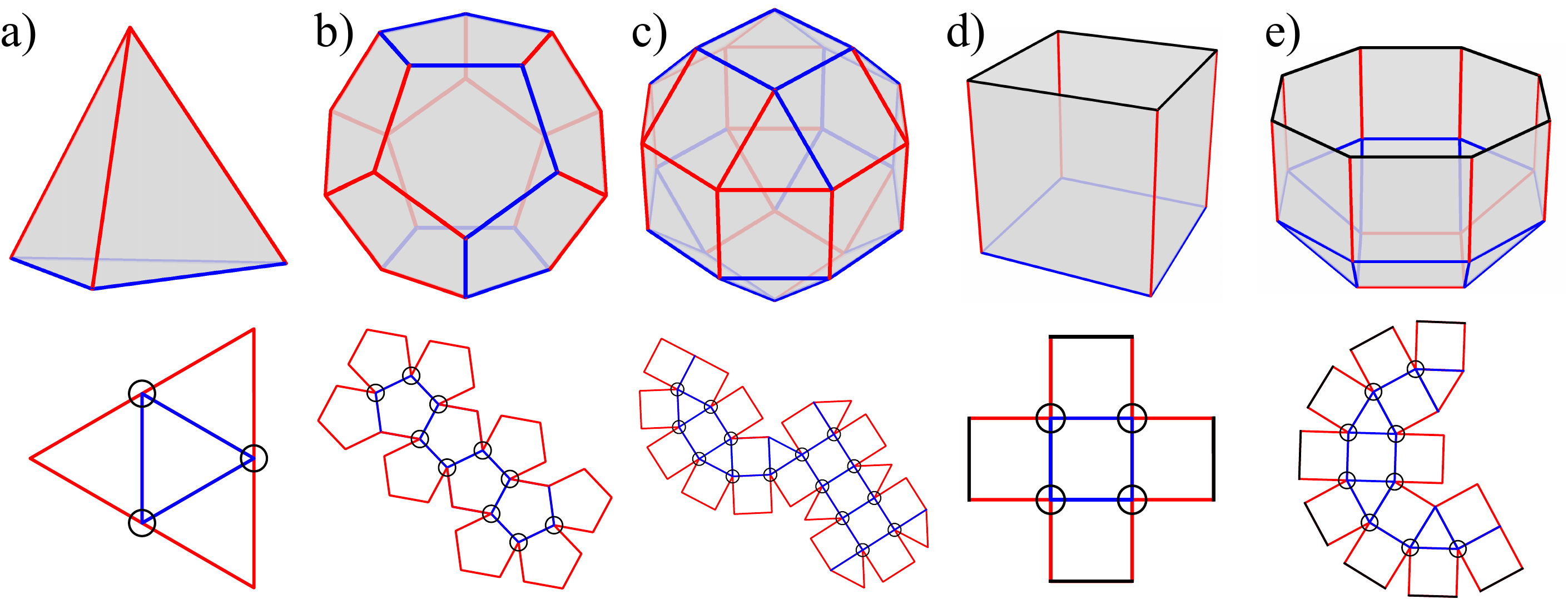}
\caption{Five examples of shells and of one of their nets corresponding to a
	cut that is a maximum leaf spanning tree: a) tetrahedron, with four
	faces and nine edges, it has four maximum leaf spanning trees, but
	only one non-isomorphic; b) dodecahedron, with twelve faces and thirty
	edges, it has $1980$ maximum leaf spanning trees, but only $21$
	non-isomorphic; c) small rhombicuboctahedron, with $26$ faces and $48$
	edges, it has $1536$ maximum leaf spanning trees, but only $32$
	non-isomorphic; d) open cubic shell, with five faces and twelve edges,
	it has only one maximum leaf spanning tree e) small
	rhombicuboctahedron with the top nine faces removed and $17$ faces,
	$36$ edges and $20$ nodes remaining, it has $720$ maximum leaf
	spanning trees, but only $90$ non-isomorphic. The black circles in the
	nets indicate the vertex connections.\label{fig::nets}}
\end{figure*}
We propose a deterministic procedure to identify the optimal net that only
requires generating a small subset of the full set of the nets of the
polyhedron.  Let us consider the case of a cubic shell. As shown in
Fig.~\ref{fig::net_cube}(a), the structure of the shell can be mapped into a
\textit{shell graph} (black nodes and links in the figure), where nodes
represent the vertices and links represent the edges. A second graph can also
be defined, the \textit{face graph}, whose nodes are the faces and links
connect pairs of adjacent faces (blue graph). Every net of the cubic shell
corresponds to a spanning tree of its face graph, i.e., a connected sub-graph
that includes all nodes but the minimum number of links (see
Fig.~\ref{fig::net_cube}(c)). The nets can be obtained by a set of splittings
along the edges of the shell graph, under the constraint that the set of nodes
in the face graph remains connected. The cut is defined as the sub-graph of
the shell graph that contains all removed shell edges (edges of the cut), as
represented in red in Figs.~\ref{fig::net_cube}(b) and (c). It consists of all
nodes and it is a spanning tree of the shell graph. The main advantage
of the mapping proposed here is that it makes possible to identify the cuts
that maximize the number of vertex connections in a systematic and
deterministic way, as explained below.

The vertex connections correspond to the nodes of degree one (leaves)
in the cut. All cuts are spanning trees of the shell graph, so cuts with the
maximum number of vertex connections correspond to maximum leaf spanning trees
(MLSTs) of the shell graph. As the total number of nodes in the spanning tree
is set, we can maximize the number of leaf nodes by minimizing the number of
nodes connected to multiple other nodes. Thus we need to find a minimal
``dominant set'' of connected nodes which traverses the shell within
``touching distance'' (one edge) of every other node. With the addition of
these edges, these other nodes are the leaves of the spanning tree, and the
distance requirement ensures that they are close enough to connect to the
dominant set.  The algorithm we use identify the MLSTs (see Supplemental
Material~\cite{SM}) is not the most efficient (see, e.g.
Ref.~\cite{Fernau2011,Fujie2003,Lucena2010}) however it finds the full list of
MLSTs while the other algorithms do not.
\begin{figure}
\centering
\includegraphics[width=\columnwidth]{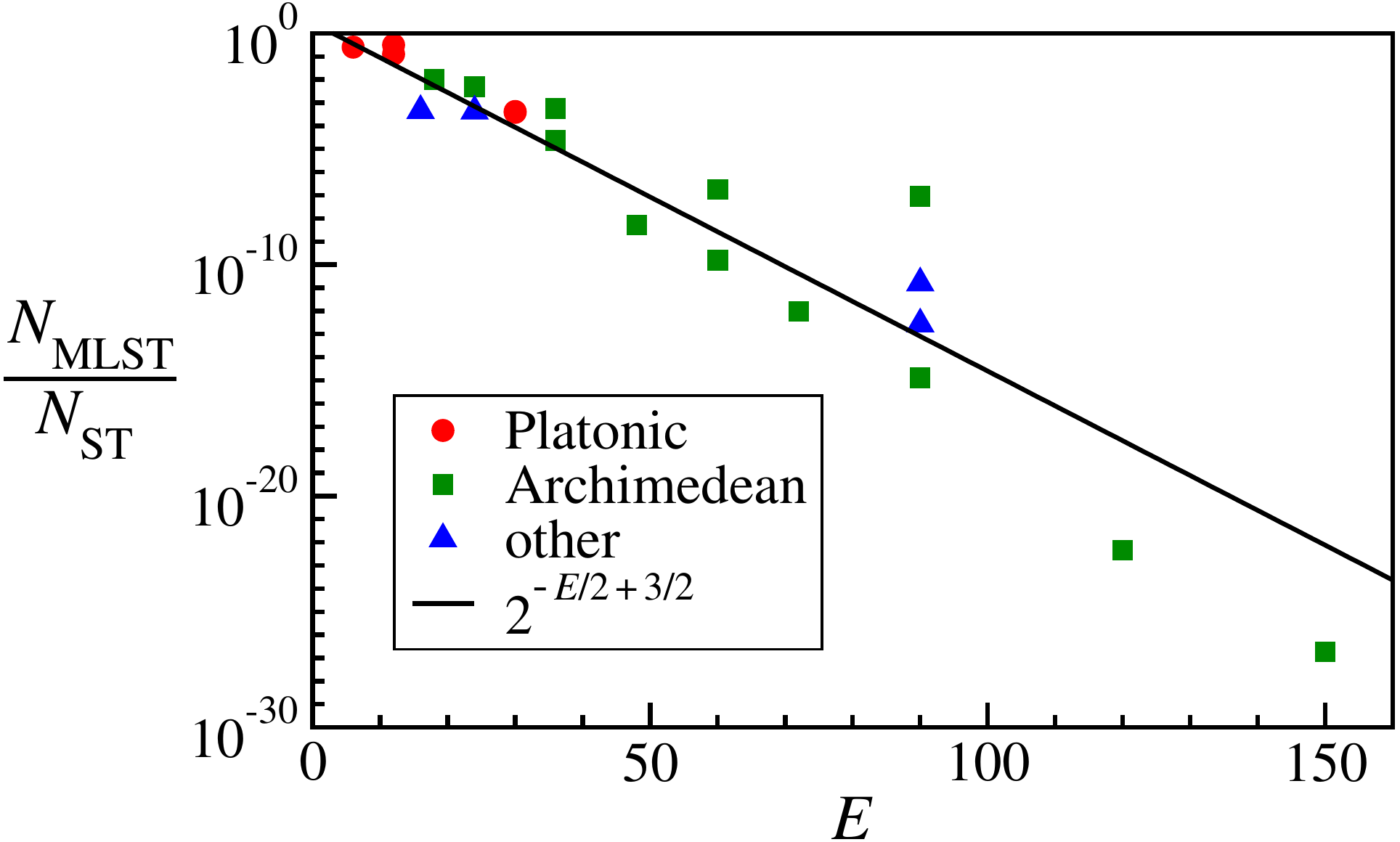}
\caption{Fraction of spanning trees ($N_\mathrm{ST}$) that are maximum leaf
	spanning trees ($N_\mathrm{MLST}$) as a function of the number of
	shell edges ($E$). This ratio was calculated for a total of $21$
	shells, including all Platonic solids and all Archimedean solids with
	up to $150$ shell edges (see Table~S1 in the Supplemental
	Material~\cite{SM}).  The solid line corresponds to the estimation
	given by Eq.~(\ref{eq::ratio.mlst}).\label{fig::fraction_spanning}}
\end{figure}

Figures~\ref{fig::nets}(a)-(c) show three examples of three-dimensional shell
structures with one of their nets, corresponding to cuts that are MLSTs. For
these examples, the number of spanning trees and MLSTs increase with the
number of shell edges (see figure caption for values).  However, the fraction
of spanning trees that are MLST decays exponentially with the number of shell
edges, $E$, as shown in Fig.~\ref{fig::fraction_spanning}. This fast decay
with $E$ reinforces the necessity of a deterministic method, since the chances
of obtaining a MLST from a random search is given by this fraction. For
example, for the dodecahedron, with only twelve faces, less than $0.04\%$ of
its more than $5$ million spanning trees are MLSTs, i.e., to obtain a MLST one
would need to randomly sample $2500$ configurations on average. For the
largest shell that we have considered ($E=150$), this number is larger than
$10^{26}$.  Thus, identifying the optimal net from random methods is
practically impossible for such large shells. To explain this exponential
decay of the fraction of spanning trees that are MLST, we first estimate how
the number of leaves, $L$, in a MLST scales with $E$. Assuming that the shell
is a convex polyhedral with regular faces and a simplified (approximated)
procedure to identify the MLST, we can predict that
\begin{equation}\label{eq::LvsE} 
L\sim E/4+2 \ \ ,
\end{equation} 
\noindent see in detail in the Supplemental Material~\cite{SM}.
Figure~\ref{fig::LvsE} shows $L$ as a function of $E$ for all Platonic and
Archimedean solids with up to $150$ shell edges. The obtained dependence is
consistent with the predicted linear dependence (solid line).  From the simple
relation~(\ref{eq::LvsE}), one can estimate an upper bound for the number of
MLSTs, $N_\mathrm{MLST}$, as a function of $E$. The exact number of spanning
trees, $N_\mathrm{ST}$, is given by the Kirchhoff's matrix-tree theorem, which
states that the total number of labeled spanning trees is given by 
\begin{equation}\label{eq::kirchhoff}
N_\mathrm{ST}=\frac{1}{V}\prod_{i=1}^{V-1}\lambda_i \ \ ,
\end{equation}
where $V$ is the number of vertices, $\lambda_n$ the eigenvalues of the
Laplacian matrix, and $\lambda_V=0$ is excluded from the product. The
Laplacian matrix $L$ of a graph with $V$ nodes is $L=D-A$, where $D$ and $A$
are the degree and adjacency matrices, respectively. The degree matrix $D$ is
a diagonal matrix where the element $D_{ii}$ is the total number of
connections of node $i$. The elements of the adjacency matrix $A_{ij}$ are
either one, if nodes $i$ and $j$ are connected, or zero otherwise.  We can get
an upper bound for $N_\text{ST}$ as a function of $E$.  Accordingly, as we
show in the Supplemental Material~\cite{SM}, an estimate for the ratio
$N_\mathrm{MLST}/N_\mathrm{ST}$ is given by
\begin{equation}\label{eq::ratio.mlst} 
N_\mathrm{MLST}/N_\mathrm{ST}\sim 2^{-E/2+3/2} \ \  , 
\end{equation}
\noindent which is also consistent over more than $20$ orders of magnitude
with the observed exponential dependence shown in
Fig.~\ref{fig::fraction_spanning}.  
\begin{figure}
\centering
\includegraphics[width=\columnwidth]{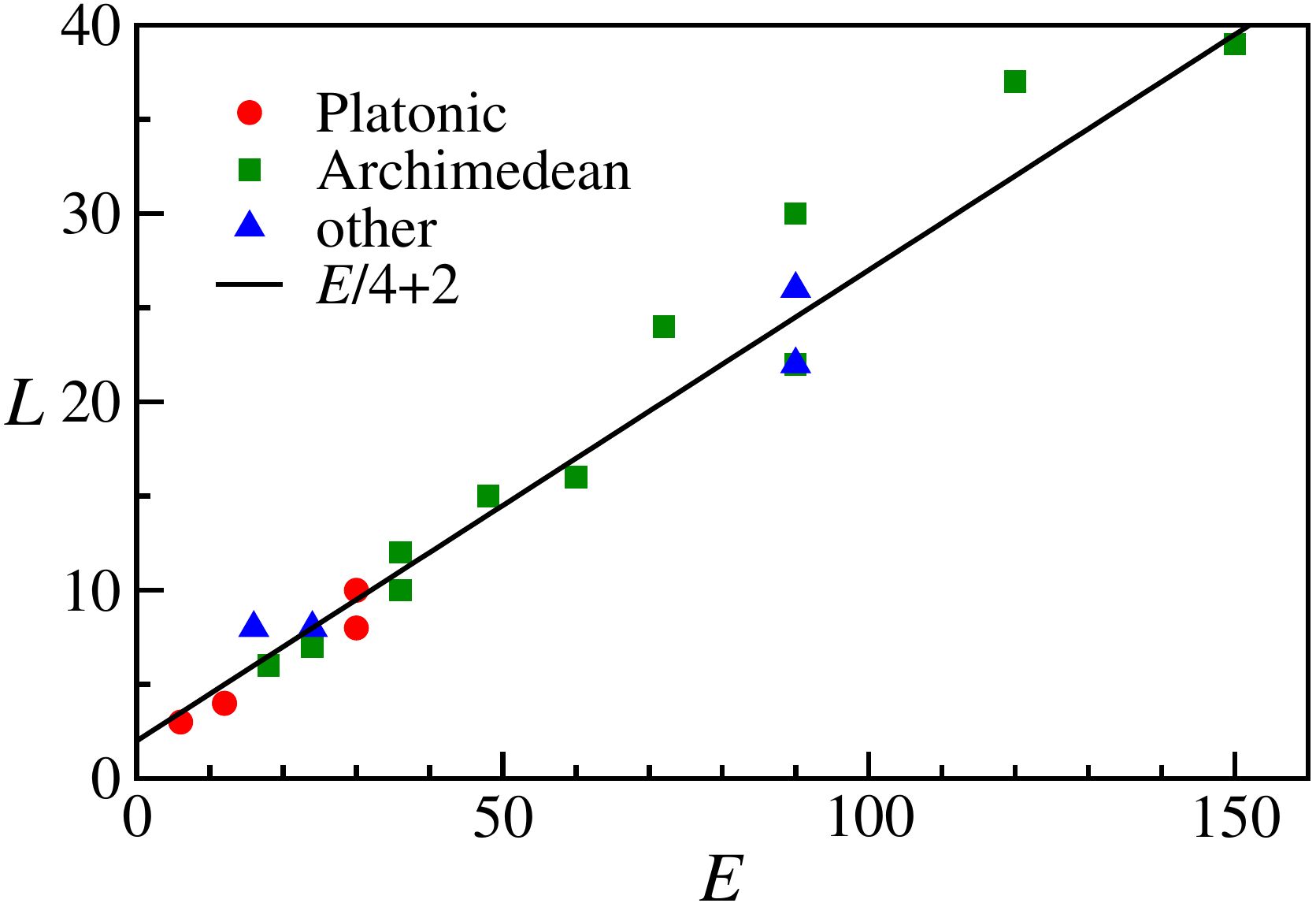}
\caption{Number of leaves ($L$) as a function of the number of shell edges
	($E$). This number was calculated for a total of $21$ shells,
	including all Platonic solids and all Archimedean solids with up to
	$150$ shell edges (see Table~S1 in the Supplemental
	Material~\cite{SM}). The solid line corresponds to the estimation
	given by Eq.~(\ref{eq::LvsE}).~\label{fig::LvsE}}
\end{figure}

With the strategy proposed here, we can also consider open structures, without
one or more faces, as the ones shown in Figs.~\ref{fig::nets}(d) and (e).
These shells might be relevant for several applications involving, for
example, encapsulation, drug delivery, or key-lock
mechanism~\cite{Azam2011,Fernandes2012,Shim2012,Filippousi2013}. The shell
graph for such structures is equivalent to the shell graph of the
corresponding (closed) polyhedron; however, every cut includes all edges
connecting two nodes of the missing face. So, to identify the optimal net in
this case, we require that the edges of the missing face are in the cut and
then follow the same procedure as before.  Note that, in this case, the cut is
no longer a tree, as the edges of the missing face form a loop, but this is
the only possible loop in the cut as any other will split the net into pieces
(see Supplemental Materials for further details~\cite{SM}).

Previous studies suggested additional criteria to identify optimal structures,
such as, the minimum radius of gyration (geometric compactness), number and
type of symmetric elements, or type and shape of intermediate structures during
the folding~\cite{Pandey2011,Azam2011}. Once we find the full set of nets with
the maximum number of so-called vertex connections, we can apply additional
criteria to a much smaller set of nets. As an example, we will consider the
additional criterion of geometric compactness where we pick the net with the
lowest radius of gyration. To apply this criterion, the label of the individual
nodes in the MLST is irrelevant and so, for simplicity, we first identify the
subset of non-isomorphic cuts, i.e., the subset of cuts that we cannot get one
from another by relabeling the nodes (or faces). This is a much smaller subset.
For example, for the cubic shell, $120$ MLSTs were identified, but only four
are non-isomorphic. To identify them, we rely on the concept of the adjacency
matrix. Two graphs are isomorphic if their adjacency matrices can be turned
into each other by a sequence of line and column swaps. Note that, each swap
corresponds to a relabeling of the nodes and thus the matching swaps need to be
done for the lines and the columns.

\begin{figure}
\centering
\includegraphics[width=\columnwidth]{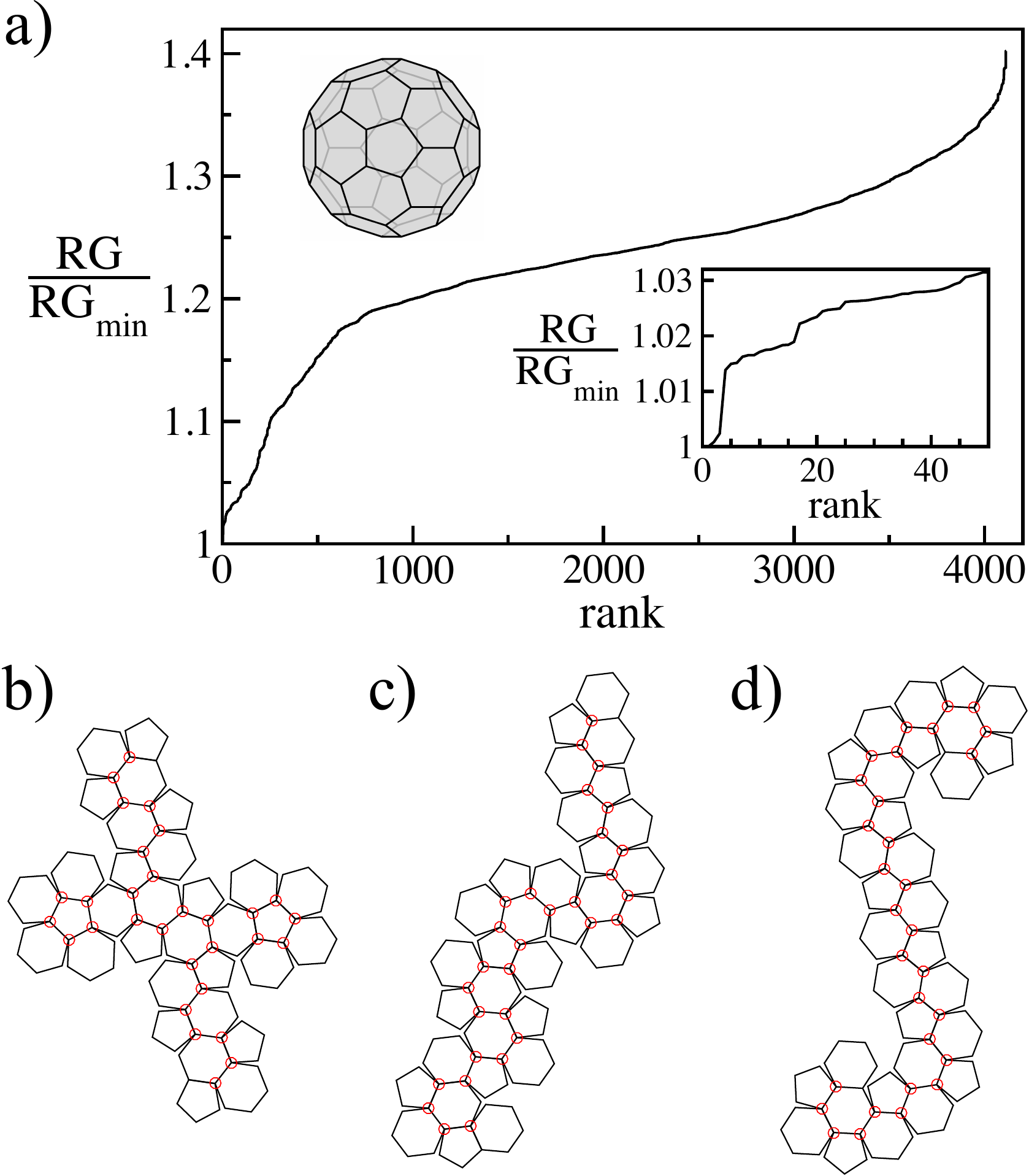}
\caption{For the truncated icosahedron, also known as soccer ball or
	buckyball: (a) spectrum of the radii of gyration for all the $4114$
	non-isomorphic MLSTs, where MLSTs were ordered by increasing radius of
	gyration; the MLST with the (b) optimal (rank 1); (c) intermediate
	(rank 2057: $\mathrm{RG}/\mathrm{RG_{min}}\approx1.23$); and (d)
	largest (rank 4114: $\mathrm{RG}/\mathrm{RG_{min}\approx1.40}$) radius
	of gyration.~\label{fig::buckyball}}
\end{figure}
Figure~\ref{fig::buckyball} shows the study for the truncated icosahedron,
also known as soccer ball or buckyball. With $60$ vertices, $32$ faces, and
$90$ edges, this shell has more than $10^{20}$ possible cuts. By mapping the
shell into a graph, we can identify the $484800$ cuts with the maximum number
of leaves ($30$ leaves each) and identify the $4114$ non-isomorphic
corresponding nets. To calculate the radius of gyration $R_g$ for each net, we
first determine the centroid of the region $\alpha\subset \mathbb{R}^2$,
corresponding to the set of all faces of the net. The centroid is defined as
\begin{equation}
(\bar{x},\bar{y}) = \frac{1}{A}\int_\alpha (x,y)\, dA \ \ ,
\end{equation}
where $A$ is the total area of the shell faces. We then calculate the radius
of gyration with respect to the centroid, defined as
\begin{equation}
	R_g = \sqrt{\frac{1}{A}\int_\alpha \left[(x-\bar{x})^2 +
	(y-\bar{y})^2\right] dA} \ \ . 
\end{equation}
To evaluate this integral, we use a standard method that only requires the
coordinates of the vertices and centroid, as explained in
Ref.~\cite{Steger96}.  Figure~\ref{fig::buckyball}(a) shows the spectrum of
the radii of gyration for all the $4114$ non-isomorphic nets of the truncated
icosahedron, ranked by increasing radius of gyration;
Figures~\ref{fig::buckyball}(b)-(c) show from those nets, the ones with the
lowest (optimal), intermediate, and maximum $R_g$. The radius of gyration
rapidly increases with the position in the rank.  The less optimal net
(Fig.~\ref{fig::buckyball}(d)) has a radius of gyration that is $40\%$ higher
than the optimal net (Fig.~\ref{fig::buckyball}(b)).  

Previous methods to identify the optimal net are based on random search. Using
such methods, even if the obtained net corresponds to one with the maximum
number of vertex connections, the probability that its radius of gyration
differs by less that $3\%$ from the optimal one is below $1.5\%$ (see inset in
Fig.~\ref{fig::buckyball}(a)).  Since the timescale and yield of the
self-folding depends strongly on the radius of gyration, the self-folding
efficiency of an approximated solution obtained with previous methods based on
a random search is likely far from optimal.

\textit{Conclusions.} We proposed a method to identify the optimal net that
spontaneously self-folds into a closed or open polyhedral shell structures.
The method consists of mapping the shell structure into a shell graph and find
the cuts that maximize the number of vertex connections.  Adapting concepts and
methods from Graph Theory, we show that the optimal solution can be obtained in
a deterministic and systematic manner. Previous methods are based on random
search and do not provide an optimal solution. As we showed, the fraction of
nets that have the maximum number of vertex connections decays exponentially
with the number of edges in the polyhedron, reinforcing the necessity of a
deterministic method.

To select the optimal net for self-folding, we considered solely the
topological and one particular measure of the geometric compactness (radius of
gyration). The authors of Ref.~\cite{Pandey2011} also discussed less-specific
features to be considered for further improvement. With the method proposed
here, additional design criteria for further compactness optimization can be
implemented straightforwardly, since a complete list of possible nets
maximizing topological compactness is obtained.

We conjectured that the nets that maximize the number of vertex
connections for all convex shells are non-overlapping, similarly to
Refs.~\cite{Shephard1975,Schlickenrieder97,Demaine07}, a necessary condition
for obtaining a shell from a two-dimensional template. For all shells
considered here, all optimal shells are non-overlapping but nets for concave
shells are more likely to overlap. In fact, this conjecture is not needed.
Once the optimal net is identified, self-overlapping can be tested.
If the selected net overlaps, one should proceed through the rank of
increasing radius of gyration and pick the first net that does not
self-overlap. If all nets for the maximum leaf spanning trees obtained 
overlap, one should proceed iteratively considering spanning trees
with less and less leaves until we found the first that does not overlap.

Identifying the maximum leaf spanning tree is a NP-complete problem and so the
numerical complexity will still grow rapidly with the number of shell
vertices.  If the number of vertices is too large for the straightforward
implementation of our deterministic algorithm then, in the spirit of our
approach, approximated algorithms can be used that identify spanning trees
with a number of leaves that is close to the maximum~\cite{lu1998,solis2017}.

We suggest that our deterministic algorithm and its variations could be used
when searching for optimal design and production of even more complex
self-assembling systems.

\begin{acknowledgments}
We acknowledge financial support from the Portuguese Foundation for Science and
Technology (FCT) under Contract no. UID/FIS/00618/2013 and grant
SFRH/BPD/123077/2016.
\end{acknowledgments}

 \FloatBarrier

%%%%%%%%%%%%%%%%%%%%%%%%%%%%%%%%%%%%%%
%%%%%%%%%%%%%%%%%%%%%%%%%%%%%%%%%%%%%%
%%%%%%%%%%  SUPPLEMENTAL MATERIAL  %%%%%%%%%%%%
%%%%%%%%%%%%%%%%%%%%%%%%%%%%%%%%%%%%%%
%%%%%%%%%%%%%%%%%%%%%%%%%%%%%%%%%%%%%%
%%%%%%%%%%%%%%%%%%%%%%%%%%%%%%%%%%%%%%

 \onecolumngrid
\clearpage
%\pagebreak

\begin{center}

\textbf{\large Finding the optimal nets for self-folding Kirigami -- Supplemental Material\\[11pt] }

{N.~A.~M.~Ara\'ujo,   R.~A.~da~Costa,   S.~N.~Dorogovtsev, and   J.~F.~F.~Mendes}
\\[20pt]

\end{center}
\twocolumngrid

\setcounter{equation}{0}
\setcounter{figure}{0}
\setcounter{table}{0}
\setcounter{page}{1}
\makeatletter
\renewcommand{\theequation}{S\arabic{equation}}
\renewcommand{\thetable}{S\arabic{table}}
\renewcommand{\thefigure}{S\arabic{figure}}
\renewcommand{\bibnumfmt}[1]{[S#1]}
\renewcommand{\citenumfont}[1]{S#1}
\renewcommand{\thesection}{S\arabic{section}} 
\renewcommand{\thepage}{S\arabic{page}}

\section{S1.\ \ \ One-to-one correspondence between nets and spanning trees}

To each net of a polyhedron corresponds a cut along the edges of a spanning
tree of the shell graph:

\noindent i) To unfold a polyhedral shell into a 2D net, the cut must reach
every vertex of the shell graph, and must be connected;

\noindent ii) For the polyhedron faces to remain connected as a single
component, the cut cannot contain any loops.

The only subgraphs that span to every vertex, are connected, and contain no
loops are spanning trees. Therefore, maximizing the number of 
vertex connections of a net is equivalent to maximizing the number of leaves of a
spanning tree.

\section{S2.\ \ \ The Maximum Leaf Spanning Tree}
\label{MLST}

The Maximum Leaf Spanning Tree (MLST) problem has been extensively studied in
the scope of graph theory and computer science
\cite{garey2002computers,rosamond2008max,fernau2011exact,fujie2003exact,lucena2010reformulations,solis20172,lu1998approximating}.
It consist of finding a spanning tree with the largest possible number of
leaves in a given undirected unweighted graph. Finding a MLST, or just
determining the number of leaves of the MLSTs of a generic graph is a well
known NP-complete problem \cite{garey2002computers}. 

Here, we describe a simple (exact) algorithm to find the full set of labeled
MLSTs of an arbitrary (undirected and unweighted) graph.  Notice that while the
algorithms for the MLST problem typically find a single optimal tree
\cite{fernau2011exact,fujie2003exact,lucena2010reformulations}, our algorithm
provides all possible labeled MLSTs. The number of spanning trees and of MLSTs
grow both exponentially with the graph size and the computation time of an
algorithm that lists all MLSTs grows at least as quickly as the number of
MLSTs. 

A dominating set of a graph is a subset of vertices, such that all vertices of
the graph either belong to the set, or are connected to at least one vertex in
the set. If, in addition, those vertices (together with the edges between them)
form a connected subgraph, this set is called a connected dominating set.
Clearly, the non-leaf vertices of any spanning tree form a connected dominating
set. Therefore, finding the maximum number of leaves in a spanning tree is
equivalent to determining the minimum size of a connected dominating set.  In
the rest of this section, a subtree of the graph whose vertices form a
connected dominating set will be called \textit{dominating subtree}. (Note that
the vertices of a subtree are connected by definition.)

\subsection{A.\ \ \ Algorithm}
In order to list all MLSTs, we use a search algorithm for finding the full set
of dominating subtrees with exactly $n_S$ vertices (i.e., subtrees with $n_S$
vertices that are a connected dominating set of the graph). We start by
checking if there are any dominating subtrees of $n_S=1$ vertices, which only
exist when a single vertex is connected to all the other vertices in the
original graph. If there is no such tree, we iteratively increase $n_S$ by $1$
and search again. The search stops when the set of dominating subtrees with
$n_S$ vertices is non-empty. These minimum dominating subtrees are the
interiors of the MLSTs, that is, the MLSTs without the leaf vertices and their
respective edges. 

To finalize the construction of the MLSTs, we attach the remaining vertices to
the obtained minimum dominating subtree. These vertices are the leaves of the
MLST. If every leaf vertex only has one edge of the original graph connecting
it to the dominating subtree, then there is only one possible MLST with that
particular interior subtree. However, some leaf vertices may have multiple
edges of the original graph linking them to the dominating subtree, and so
there are multiple MLSTs with the same interior subtree. In these cases, we
have to chose one of the possibilities for each leaf vertex. Since those
choices are independent from each other, the total number of MLSTs that share
that particular interior subtree equals the product of the numbers of
possibilities of each leaf vertex (i.e., the number of different ways of
connecting the leaf vertices to a particular dominating subtree).

The algorithm recursively grows subtrees with $n_S$ vertices and $n_S-1$ edges
of the graph starting from a single root vertex and enumerates all MLSTs where
the root is a non-leaf vertex. There is a set of vertices, say $R$, that the
algorithm uses as roots (one at each time), and consists of a specific
arbitrary vertex and all of its neighbors. For any dominating subtree, all the
vertices of the original graph must either be in the subtree or have at least
one neighbor in the subtree. To find all MLSTs, it is sufficient to consider
the roots in $R$. In the present work, we use the vertex with the smallest
degree and its neighbors as the set of roots $R$.

Given a size of the subtrees $n_S$, the algorithm performs a separate search
for each root vertex in $R$. So, we need an additional constraint to avoid
multiple counts of the MLSTs that include more than one non-leaf vertices in $R$.
We introduce a set of vertices, $V_\text{excl}$, that are explicitly forbidden
from joining the dominating subtree. In the first search, rooted in the first
vertex of $R$, say $r_1$, the set $V_\text{excl}$ is empty and the algorithm
enumerates all MLSTs where $r_1$ is a non-leaf vertex.  Since the vertex $r_1$
must be a leaf in all the other MLSTs still not found, we add $r_1$ to
$V_\text{excl}$.  In the second search, rooted in $r_2$, the vertex $r_1$ is
never included in dominating subtree, insuring that the algorithm only returns
the MLSTs where $r_1$ is a leaf and $r_2$ is a non-leaf vertex.  Then, $r_2$ is
added to $V_\text{excl}$, the third search is made, and so on.

Given a graph with $V$ vertices labeled $i=1,...,V$, let us denote the edge
that connects the vertices $i$ and $j$ by $e_{ij}$. In the following
Algorithms~\ref{main}-\ref{hole} the set of vertices that are connected to $i$
in the original graph is denoted $A_i$, and the set $A=\{A_i: i=1,..,V\}$
provides complete information about the graph. 

\begin{algorithm}[H]
\begin{algorithmic}
\caption{\textit{Enumeration of all MLSTs.} This procedure initializes the necessary variables, and calls the function \text{\textsc{recursive}} of Algorithm~\ref{recursive}. 
Each time it is called, \text{\textsc{recursive}} enumerates all MLSTs with $L-n_S$ leaves, where the supplied root, $r$, in a non-leaf vertex and the vertices in $V_\text{excl}$ are leaves.
}
\label{main}
\Procedure{list\_all\_MLSTs}{$A$}
\\
\textbf{Input:}\\
List of neighbors, $A_i$, for every vertex $i$ ($A=\{A_i\}$).\\

	\State $v=\text{arbitrary vertex}$ \Comment{For example, a vertex with the minimum degree}
	\State $R=\text{append}(v,A_v)$
	\State $MLSTs=\emptyset$
	\State $n_S=1$
	\While{$MLSTs=\emptyset$}
	
		\State $V_\text{excl}=\emptyset$
	
		\ForAll{ $r \in R$ }
			\State $E_A=\{e_{ri}: i\in A_r\}$
			\State ${MLSTs}' {=} \Call{recursive} {A,\! \{r\},\! \emptyset,\! E_A,\! V_\text{excl},\! n_S}$
			\State $MLSTs=\text{append}(MLSTs,{MLSTs}')$

			\State $V_\text{excl}=\text{append}(V_\text{excl},r)$ 

		\EndFor
		\State $n_S=n_S+1$
	\EndWhile

\EndProcedure
\end{algorithmic}
\end{algorithm}

The procedure \text{\textsc{list\_all\_MLSTs}}, shown in Algorithm~\ref{main},
initializes the set of roots $R$ as described above, and the current size of
the searched subtrees, $n_S$, is initialized to $1$.  The set $MLSTs$ will
store the collection of found MLSTs, and starts as an empty set. The search of
MLSTs with larger $n_S$, and a smaller number of leaves $L=V-n_S$, will only
proceed if the set $MLSTs$ remains empty.  The list of vertices $V_\text{excl}$
stores the roots of $R$ that were already used for each particular value of
$n_S$, and is initialized as an empty set every time that $n_S$ is incremented.
The search itself, is performed by the recursive function
\text{\textsc{recursive}}, Algorithm~\ref{recursive}. For each considered
$n_S$, the function \text{\textsc{recursive}} is called by the procedure
\text{\textsc{list\_all\_MLSTs}} once for each root in $R$.

The function \text{\textsc{recursive}} of Algorithm~\ref{recursive} starts with
a single vertex (root) and recursively grows subtrees $T$ up to a predetermined
size ($n_S$ vertices), while keeping track of the elements already in the tree
and at its border.  Let $V_T$ and $E_T$ be the lists of vertices and edges,
respectively, currently in $T$, and $E_A$ be the list of edges that are not in
$T$ and are connected to at least one vertex in $T$ ($E_A$ is the exterior
boarder of $T$, which the algorithm is currently exploring). Furthermore,
$V_\text{excl}$ is a specific set of vertices that are forbidden to participate
in the subtree $T$ (these vertices are the roots of the previous searches for
the same $n_S$).  When $T$ reaches the target size ($n_S$ vertices) it stops
increasing. If the vertices $V_T$ form a dominating set, then the algorithm
enumerates the possible ways of joining each vertex outside of $V_T$ to one and
only one vertex in $V_T$ by an edge. Each different way of making these last
connections represents a different (labeled) spanning tree whose leaves are the
vertices outside of $V_T$.

In the first stage, while the number of vertices in the growing subtree
$|V_T|<n_S$, the function \text{\textsc{recursive}} considers all possibilities
for the next edge addition to the subtree $T$ from the set of adjacent edges
$E_A$.  For each of those possibilities, \text{\textsc{recursive}} is called
again with the updated ${V_T}'$, ${E_T}'$, and ${E_A}'$.  Note that, when an
edge in $E_A$ connects two vertices already in $V_T$ it cannot be added to the
tree (because it would close a loop in $T$).  In order to keep the vertices in
the set $V_\text{excl}$ outside of $T$, the edges that lead to those vertices
are not added to $E_A$ in the update. In this way, \text{\textsc{recursive}}
finds all configurations of subtrees with $n_S$ vertices that include the root
and exclude all the vertices in $V_\text{excl}$. In the second stage, when
finally $|V_T|=n_S$, \text{\textsc{recursive}} checks if the vertices in $V_T$
form a dominating set: if so, it finishes the construction of the spanning
trees by connecting the leaves in every possible way, otherwise it returns an
empty set.  

In the search algorithm presented here, we consider only connected sets of
vertices (subtrees), and check if they are dominating. Furthermore, we only
consider subtrees that include either a specific arbitrary vertex or at least
one of its neighbors (if a subtree is dominating, every vertex of the graph
fulfills this requirement).  The combination of these two strategies
drastically reduces the configurational space and computation time.

\begin{algorithm}[H]
\begin{algorithmic}
\caption{\textit{Recursive search function.} 
This function generates all possible subtrees $T$ with $n_S$ vertices that include the root vertex and exclude all the vertices in $V_\text{excl}$. If $T$ is a dominating subtree, then the function lists all possible spanning trees that can be obtained by joining the remaining vertices of the graph to $T$.
}
\label{recursive}

\Function{recursive}{$A$, $V_T$, $E_T$, $E_A$, $V_\text{excl}$, $n_S$}
\\
\textbf{Input:}\\
List of neighbors, $A_i$, for every vertex $i$ ($A=\{A_i\}$).\\
Lists of vertices, $V_T$, and edges, $E_T$, currently in $T$.\\
List of edges, $E_A$, currently adjacent to $T$.\\
List of vertices, $V_\text{excl}$, excluded from $T$.\\
Final number of vertices, $n_S$, in $T$.\\
\textbf{Output:} \\
List of spanning trees, $STs$, with at least $V-n_S$ leaves (each spanning tree is returned as the list of its edges).
\\
\If{$|V_T|$ $<$ $n_S$} 
\Comment{Add one edge to $T$.}
	
	\State $STs=\emptyset$
		
	\ForAll{$e_{jk} \in E_A$}	
		
		\State $E_A = E_A \backslash \{e_{jk}\}$
		
		\If{ $k \notin V_T$}
			\State $i=k$
		\ElsIf{ $j \notin V_T$}
			\State $i=j$
		\Else
			\State \textbf{continue}
		\EndIf

		\State ${E_T}' =  \text{append}(E_T, e_{jk})$
		\State ${V_T}' =  \text{append}(V_T, i)$
		
		\State ${E_A}'=E_A$
		\ForAll{$l \in A_i$}
			\If{ $e_{il} \notin E_T \textbf{ and }  l\notin V_\text{excl}$}
				\State ${E_A}' = \text{append}\left( {E_A}', e_{il} \right)$
			\EndIf
		\EndFor
		
		\State ${STs}' {=} \text{ \Call{recursive}{$A$,\! ${V_T}'$\!,\! ${E_T}'$\!,\! ${E_A}'$\!,\! $V_\text{excl}$,\! $n_S$}}$
		\State $STs = \text{ append}(STs,{STs}')$
		
	\EndFor
	
	\Return $STs$

\ElsIf{$V_T$ is a \textit{dominating set}} 
\Comment{Expand $T$ into corresponding spanning trees.}
	\State $STs= \{E_T\} $
	\ForAll{vertices $i \notin V_T$}
		\State ${STs}'=\emptyset$
		\ForAll{vertices $j\in A_i \cap V_T$}
			\ForAll{trees $U \in STs$}
				
				\State ${STs}'{=}\text{append}\left({STs}',  \text{append}(U , e_{ij})\right)$  
				
			\EndFor	
		\EndFor
		\State $STs={STs}'$
	\EndFor

	\Return $STs$

\Else

	\Return $\emptyset$

\EndIf

\EndFunction

\end{algorithmic}
\end{algorithm}

The order by which the edges of $E_A$ are picked in the outermost for-cycle of
the search of \text{\textsc{recursive}} is not specified in
Algorithm~\ref{recursive} because any order will produce the same result.  In
our implementation of the function \text{\textsc{recursive}}, we picked the
edges in $E_A$ using a first-in-first-out method and effectively performed a
breadth-first search. If, for instance, the edges of $E_A$ are picked in a
last-in-first-out fashion, then the algorithm will perform depth-first search
instead.

The algorithm described in this section for listing all MLSTs is defined for
labeled graphs, and finds the set of labeled MLSTs.  If the original graph has
automorphisms, i.e., a relabeling that results in the same labeled graph, then
the set of labeled MLSTs may have multiple `copies' of the same unlabeled MLST
with different labelings (isomorphic MLSTs).  In this work, we consider the nets
of polyhedra with high regularity, such as Platonic and Archimedean solids
where all vertices are equivalent.  The polyhedral graphs of polyhedra where
all vertices are equivalent, or with sets of equivalent vertices, have
automorphisms.  However, each geometrically distinct net is entirely determined
by a cut of the polyhedral graph along the edges of an unlabeled spanning tree.
So, to determine the number of distinct nets, we need to disregard the labels
of MLSTs, and search for isomorphisms in the set of MLSTs found by the
algorithm. To check if any two labeled MLSTs are isomorphic, we simply check if
there is an automorphic relabeling of the original graph that maps one labeled
MLST into the other, see Section~\ref{auto} for details.

Typically the algorithms for the MLST problem are designed to find the number
of leaves in a MLST, and not the full set of MLSTs
\cite{rosamond2008max,fujie2003exact,fernau2011exact,lucena2010reformulations}.
Some of these algorithms include multiple stages of optimization, which include
heuristics, the use of approximated algorithms to make initial guesses, and
other sophisticated approaches. Due to the complexity of this problem, and to
the best of our knowledge, the fastest existing algorithms can find one MLST
only in graphs with up to roughly $200$ vertices
\cite{lucena2010reformulations}.

In this work, we find not just one MLST, but we list also the full set of
optimal nets (unlabeled MLSTs) for polyhedral graphs with up to $60$ vertices.
The number of different optimal nets strongly depends on the details of the
polyhedral graph.  For instance, the Truncated Icosahedron and the Truncated
Dodecahedron each have $60$ vertices, $32$ edges, and $90$ faces, however their
numbers of optimal nets are $4\ 114$ and $3\ 719\ 677\ 167$, respectively.
Table~\ref{t2} summarizes the exact results obtained in this work with the
algorithm presented here, namely the number of leaves in a MLST and the number
of optimal nets of each of the $21$ polyhedron considered. Table~\ref{t2} also
shows a figure of the optimal net which minimizes the radius of gyration for
each polyhedra.

Finally, it should be mentioned that there are several approximated algorithms
that find spanning trees with a high number of leaves, close to the maximum
possible.  If a polyhedral graph is too large to solve by exact methods, these
approximated algorithms can find a near-optimal solution in linear or almost
linear time \cite{solis20172,lu1998approximating}.

\begin{table}
\begin{tabular}{  >{\centering\arraybackslash}m{2.6cm} | >{\centering\arraybackslash}m{0.34cm} | >{\centering\arraybackslash}m{0.34cm} | >{\centering\arraybackslash}m{0.5cm} | >{\centering\arraybackslash}m{0.34cm} | >{\centering\arraybackslash}m{1.9cm} | >{\centering\arraybackslash}m{1.6cm} }
\centering
  & $V$ & $F$ & $E$ & $L$ &$N_\text{opt nets}$ &  Min RG \\[2pt]
     \hline  &&&&&& \\[-10.1pt]
Tetrahedron 				& 	$4$ & $6$ &   $4$ & $3$ & $1$ & \raisebox{-2.2pt}{\includegraphics[scale=0.05]{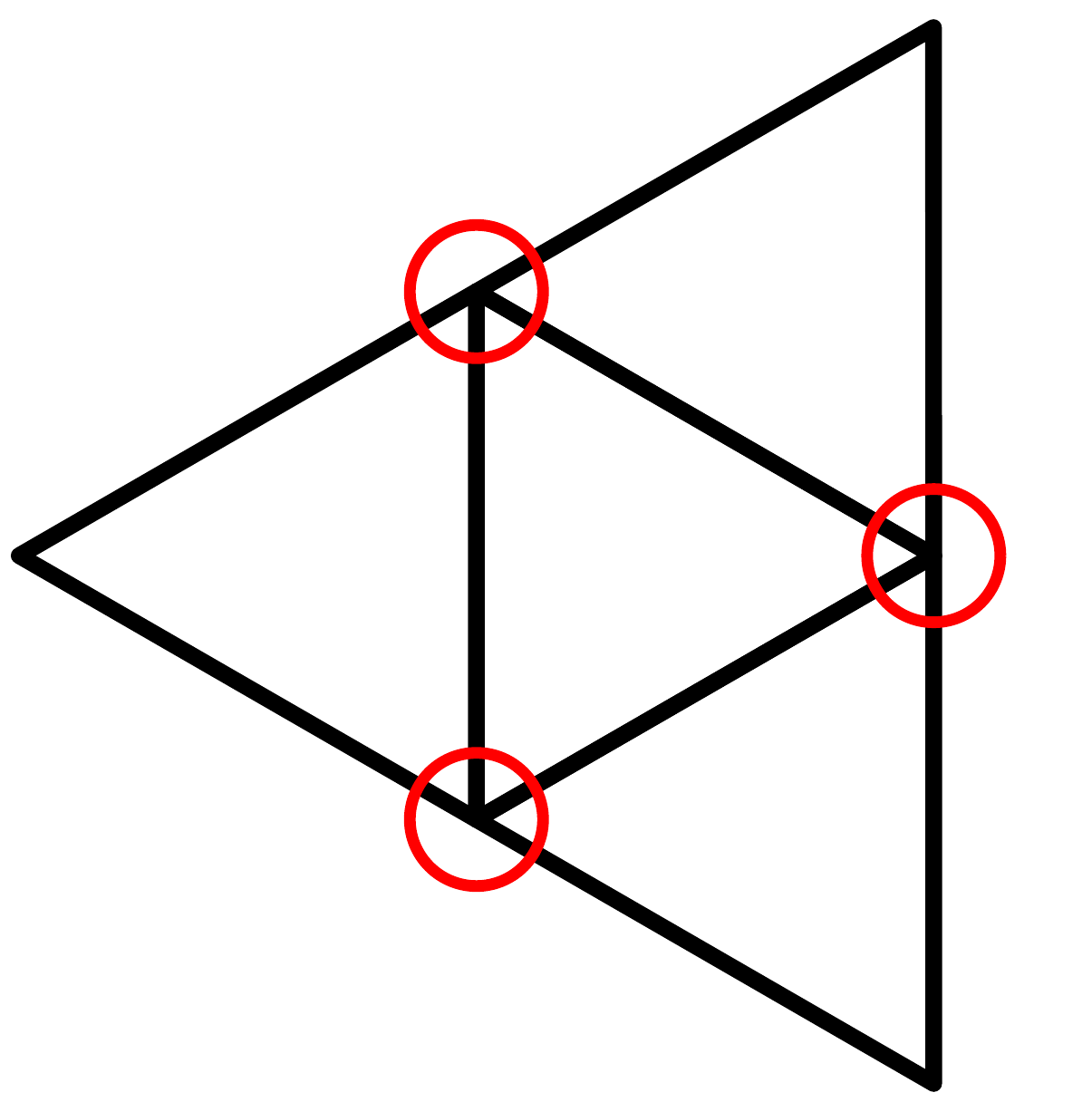}}
\\ \hline  &&&&&& \\[-10.1pt]
Octahedron                  		&	$6$ & $8$ & $12$ & $4$ & $2$ &  \raisebox{-2.2pt}{\includegraphics[angle=90,scale=0.08]{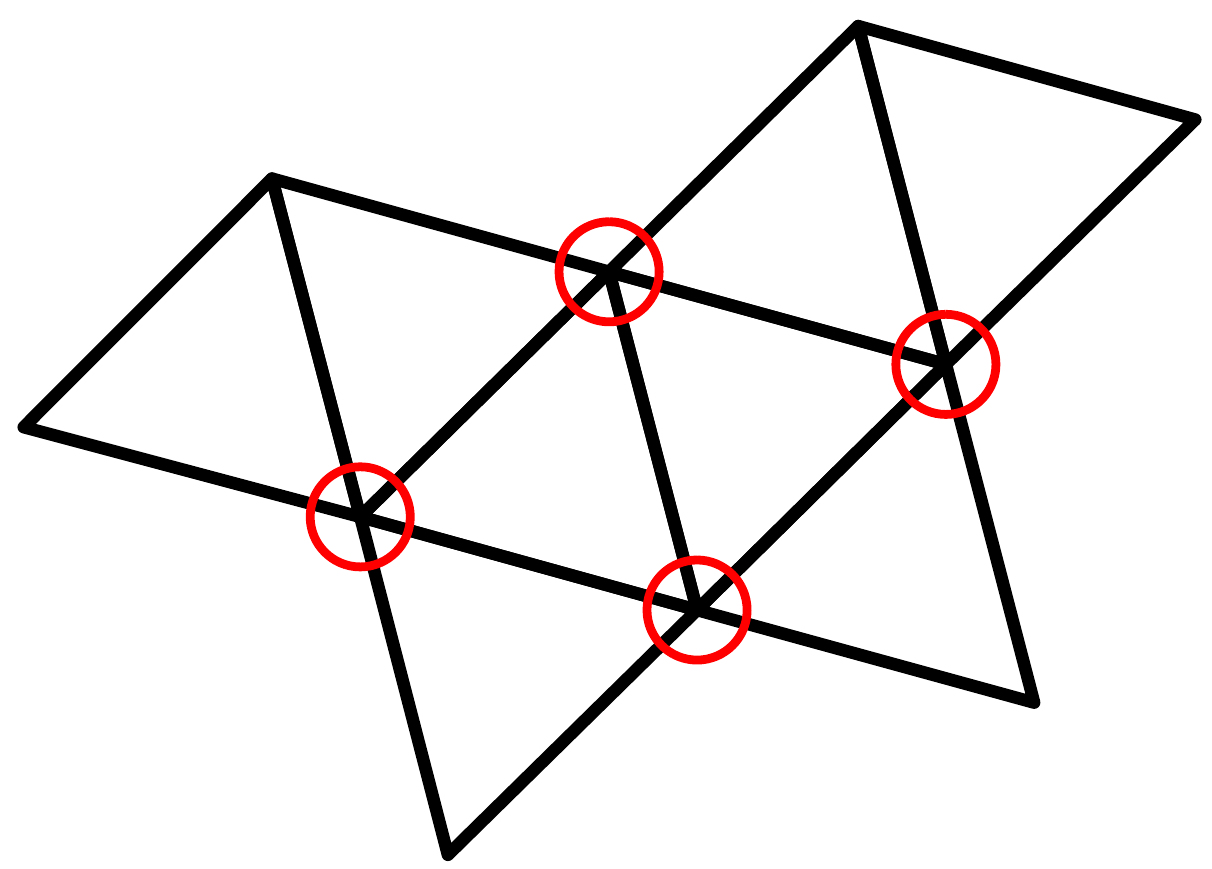}}
\\ \hline  &&&&&& \\[-10.2pt]
Cube                           	 	&	$8$ & $6$ & $12$ & $4$ & $4$ &  \raisebox{-2.3pt}{\includegraphics[scale=0.09]{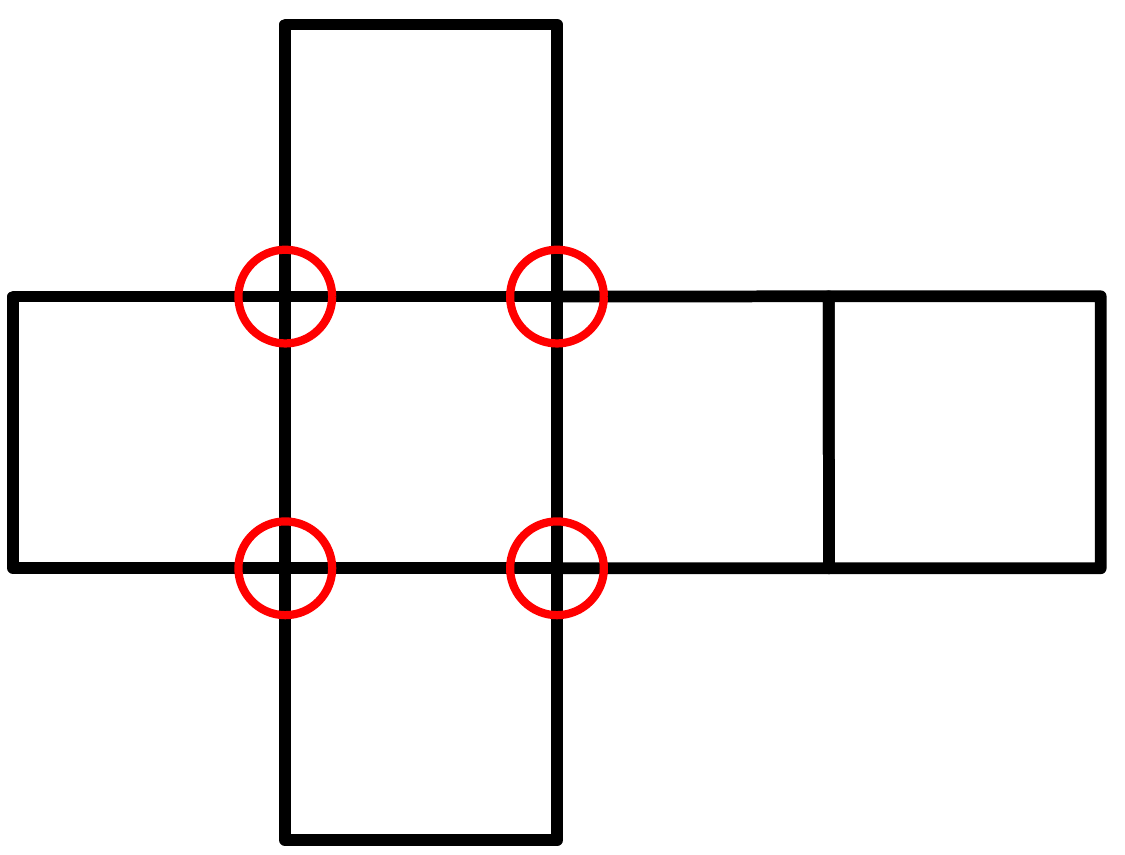}}
\\ \hline  &&&&&& \\[-10.2pt]
Icosahedron                    		&	$12$ & $20$ & $30$ & $8$ & $21$ & \raisebox{-2.2pt}{\includegraphics[scale=0.12]{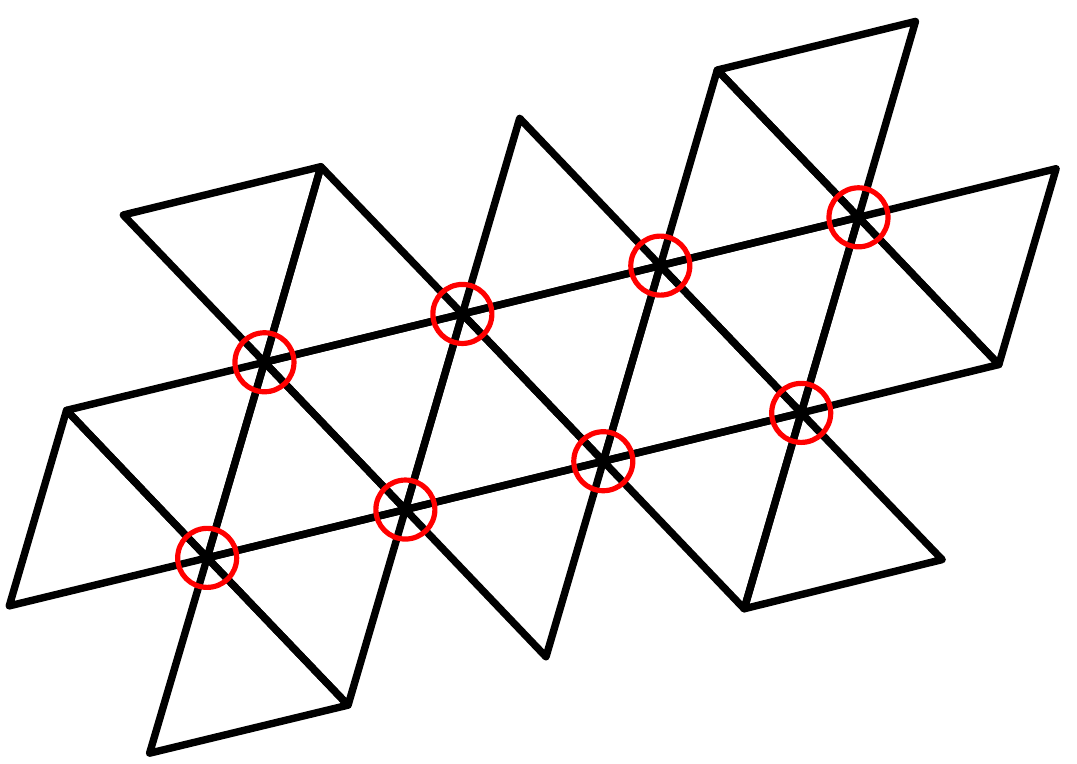}}
\\ \hline &&&&&& \\[-10.3pt]
Dodecahedron                   	&	$20$ &$12$ & $30$ & $10$ & $21$ & \raisebox{-2.2pt}{\includegraphics[scale=0.135]{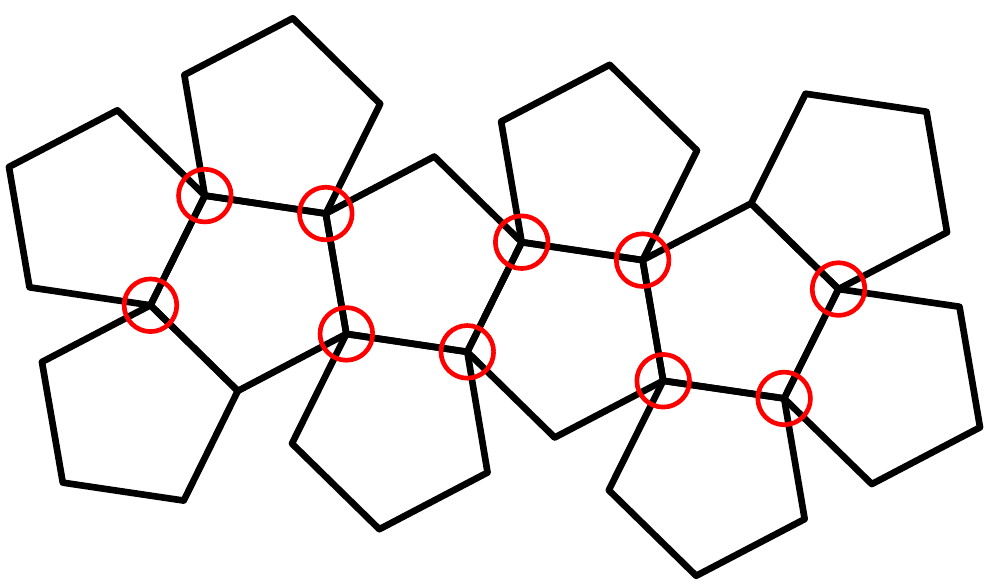}}
\\ \hline &&&&&& \\[-10.1pt]
\makecell{Octogonal \\Piramid}                	&	$9$ & $9$ & $16$ & $8$ & $1$	& \raisebox{-2.3pt}{\includegraphics[scale=0.07]{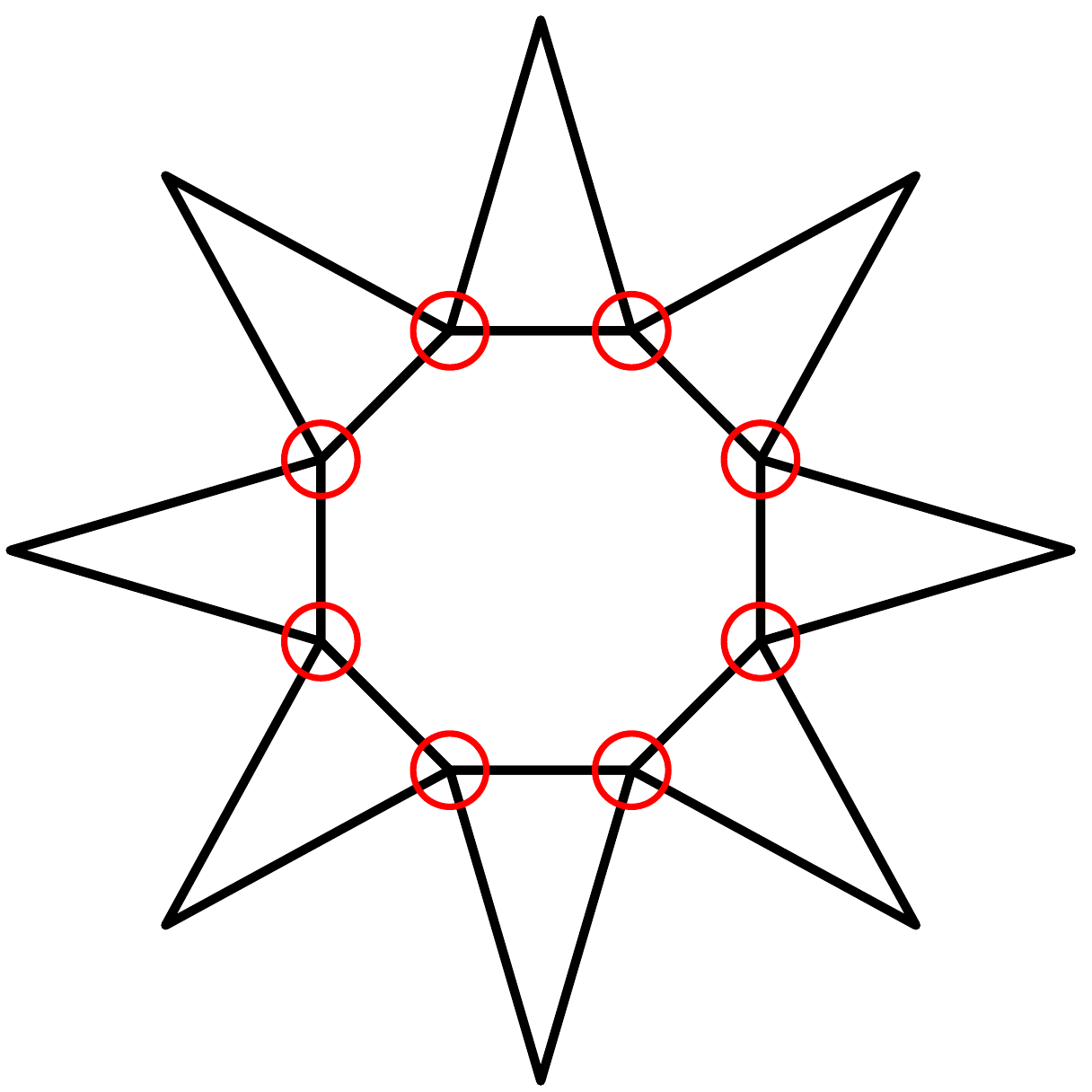}}
\\ \hline &&&&&& \\[-10.1pt]
\makecell{Octogonal \\Dipiramid}            	&	$10$& $16$ & $24$ & $7$ & $3$ & \raisebox{-2.3pt}{\includegraphics[scale=0.08]{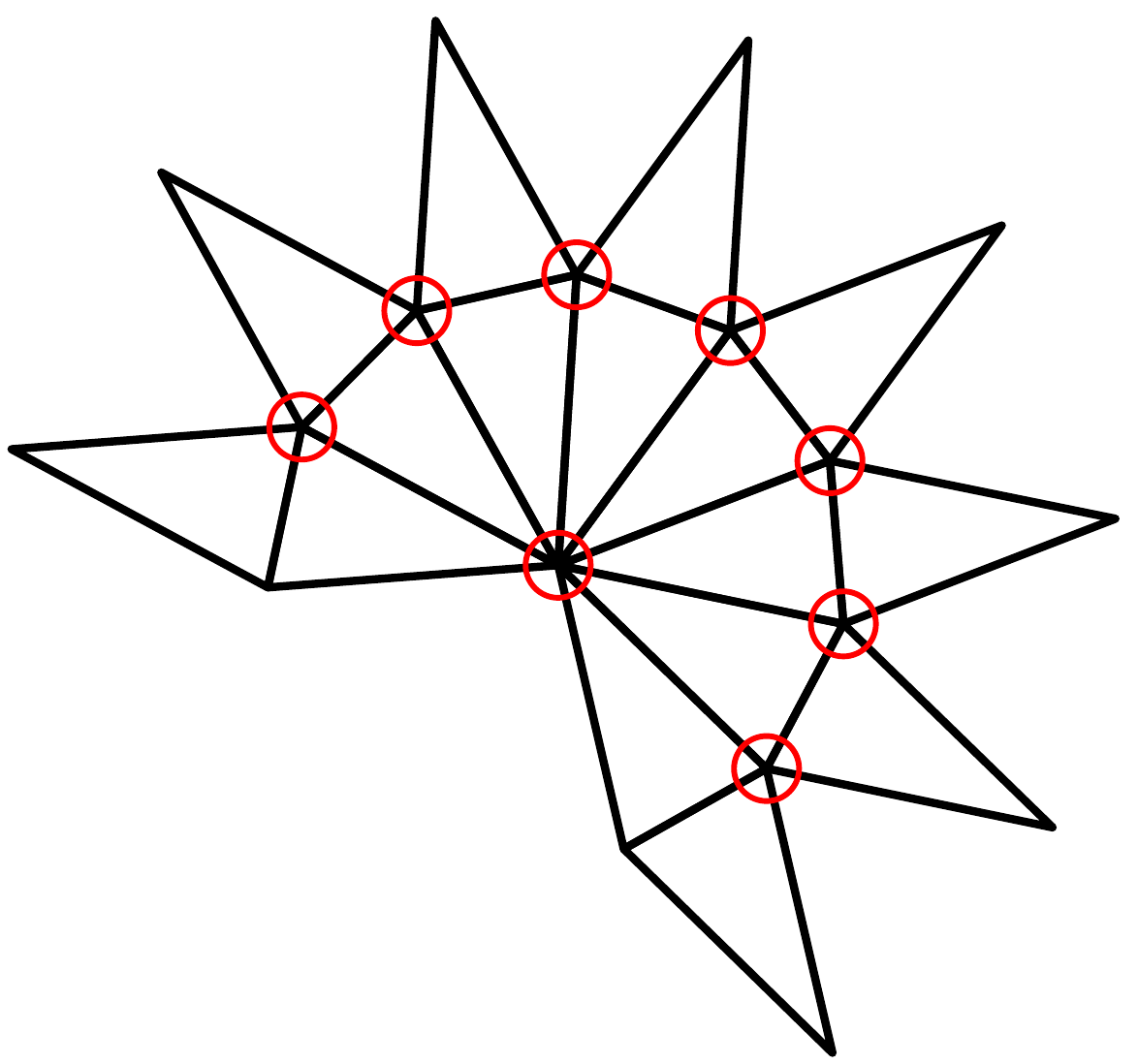}}
\\ \hline &&&&&& \\[-10.2pt]		 
\makecell{Truncated\\ Tetrahedron}          	&	$12$ & $8$ & $18$ & $6$ & $4$ & \raisebox{-2.2pt}{\includegraphics[scale=0.1]{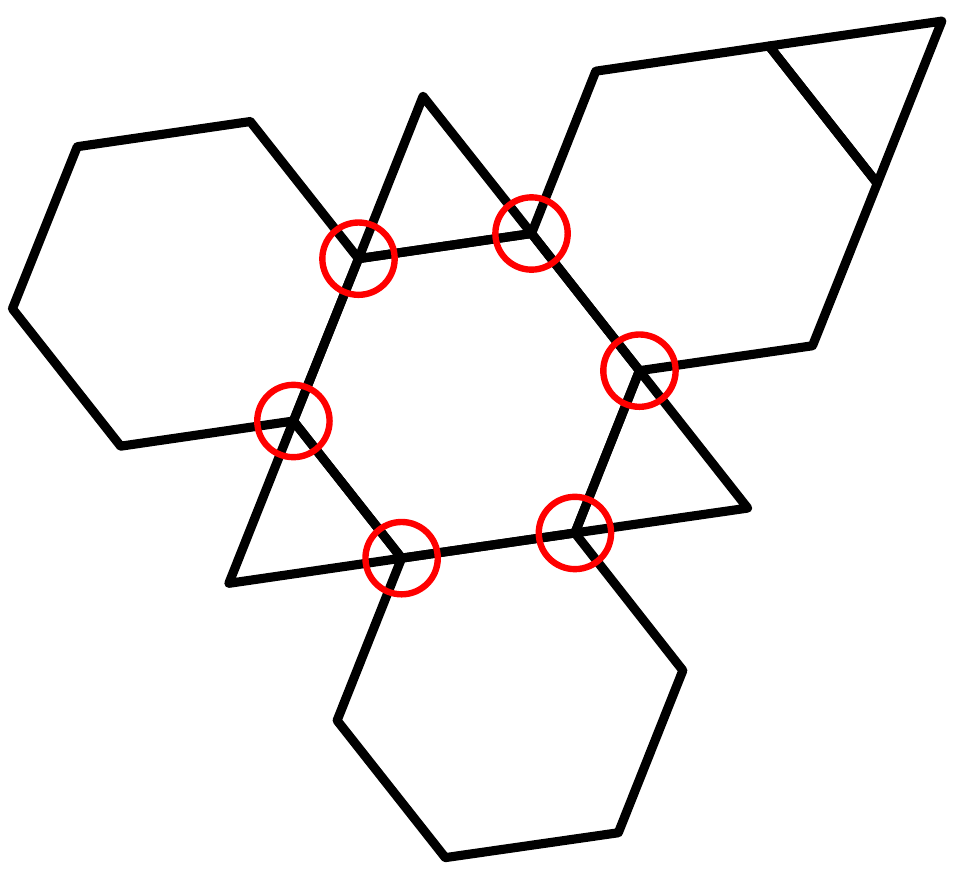}}
\\ \hline &&&&&& \\[-10.2pt]
Cuboctahedron                  	&	$12$& $14$ & $24$ & $7$ & $34$ & \raisebox{-2.2pt}{\includegraphics[scale=0.11]{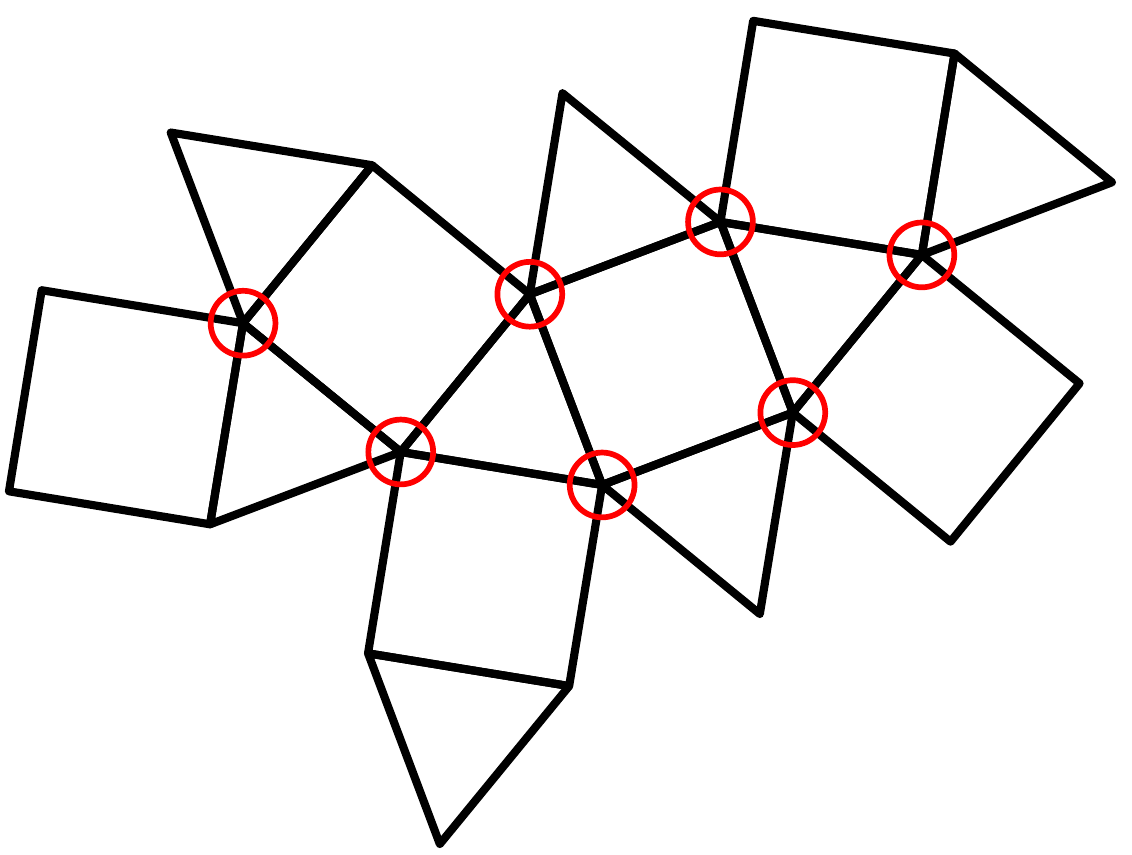}}
\\ \hline &&&&&& \\[-10.3pt]
Truncated Cube               		& 	$24$& $14$ & $36$ & $10$ & $399$	 & \raisebox{-2.2pt}{\includegraphics[scale=0.16,angle=90]{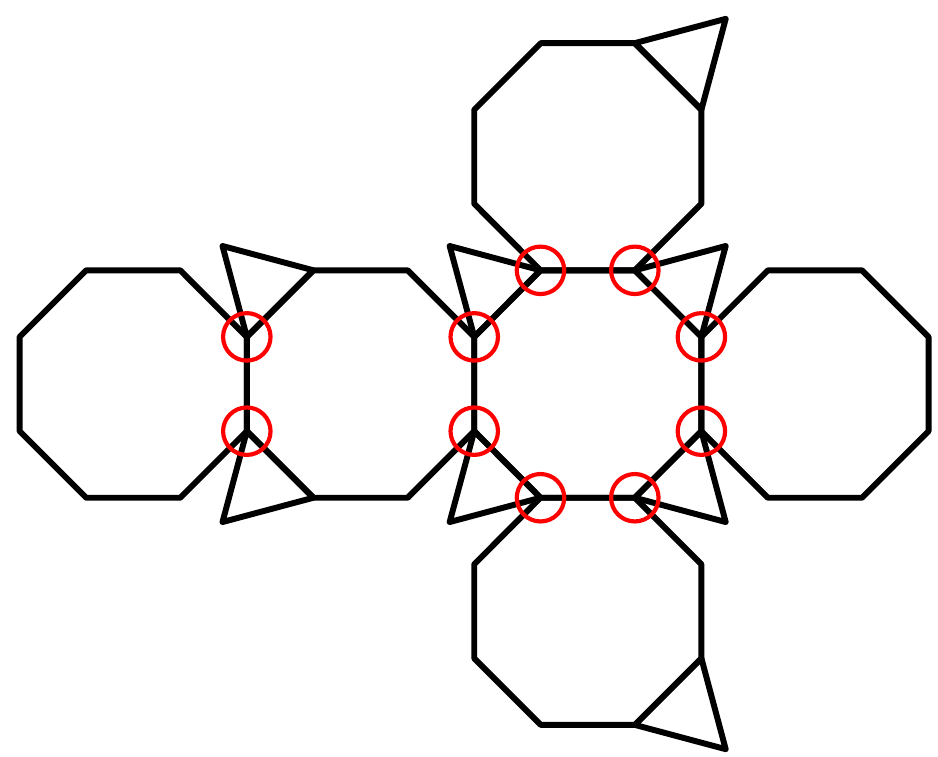}}
\\ \hline	&&&&&& \\[-10.3pt]
Snub Cube                      		&	$24$& $38$ & $60$ & $16$ & $600$	 & \raisebox{-2.2pt}{\!\includegraphics[scale=0.15,angle=90]{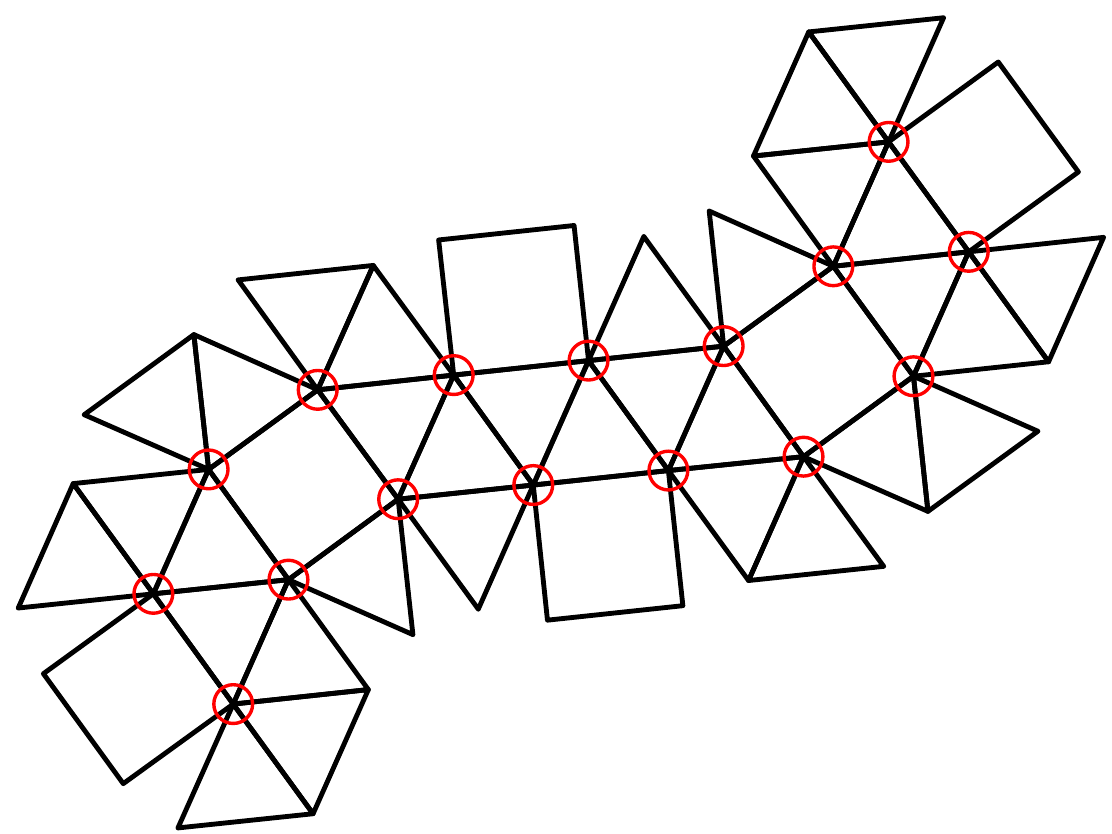}}
\\ \hline&&&&&& \\[-10.3pt]
% Small Rhombicuboctahedron 
\makecell{Rhombicu-\\boctahedron}   &  	$24$& $26$ & $48$ & $15$ & $32$ & \raisebox{-2.2pt}{\!\includegraphics[scale=0.155,angle=90]{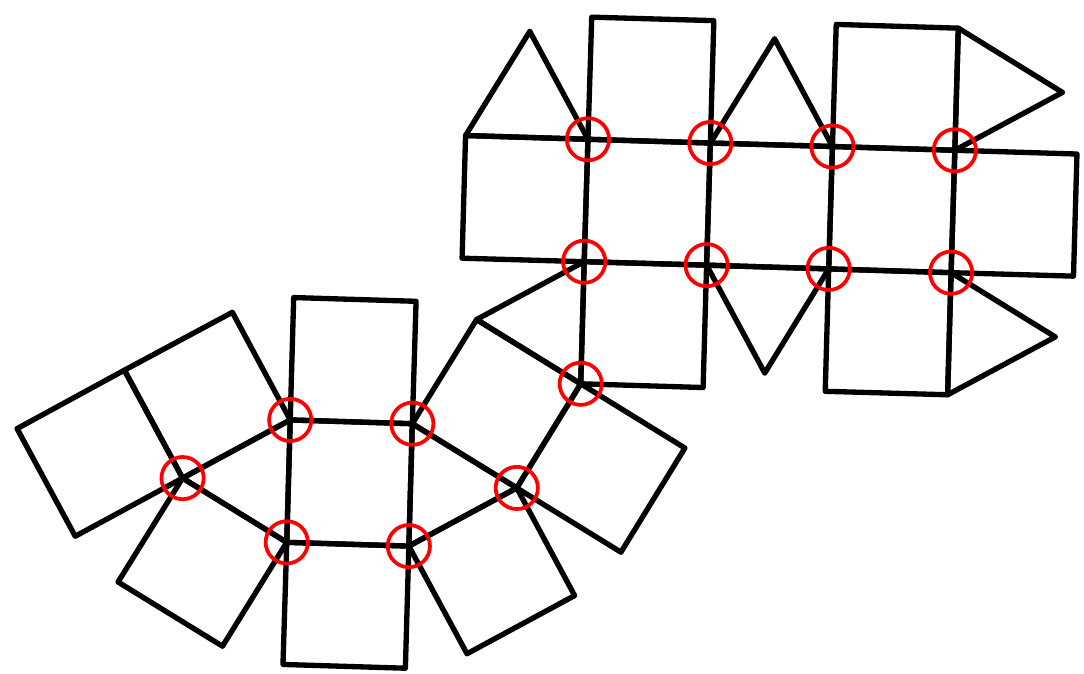}}
\\ \hline &&&&&& \\[-10.3pt]
\makecell{Truncated \\Octahedron}           	&	$24$& $14$ & $36$ & $12$ & $56$ & \raisebox{-2.2pt}{\includegraphics[scale=0.135]{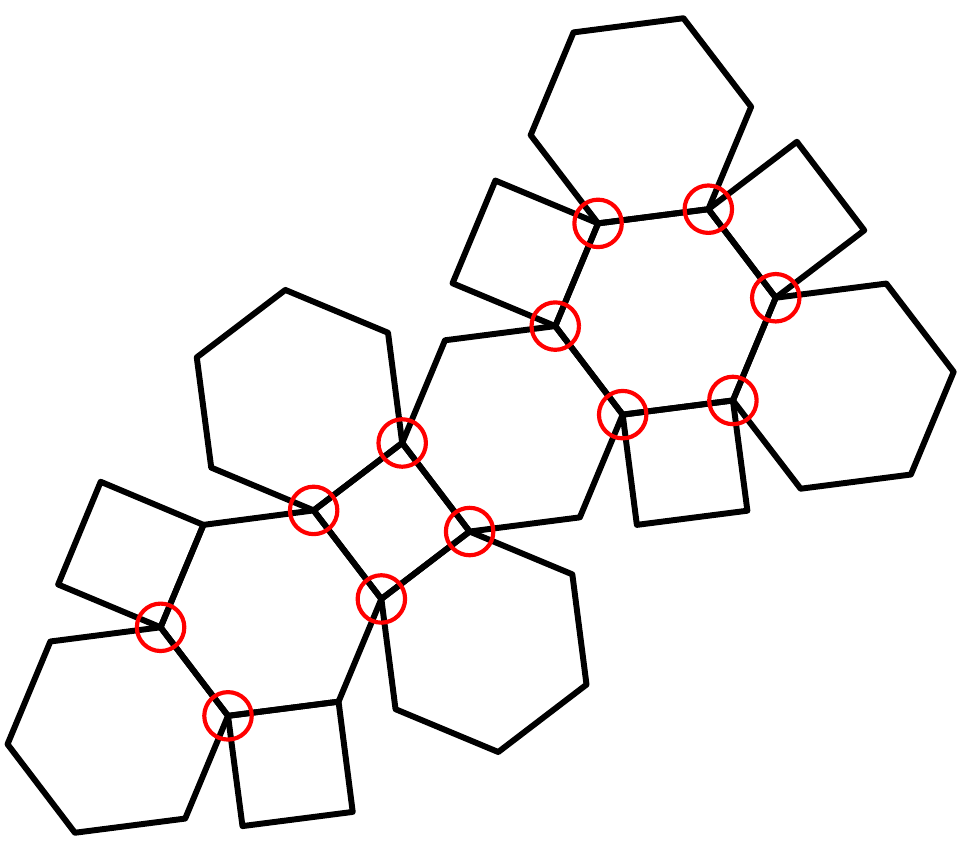}}
\\ \hline &&&&&& \\[-9.6pt]
Icosidodecahedron             	&	$30$& $32$ & $60$ & $16$ & $308\ 928$  & --
\\ \hline &&&&&& \\[-10.5pt]
%Great Rhombicuboctahedron
\makecell{Truncated\\ Cuboctahedron}    &  	$48$& $26$ & $72$ & $24$ & $244$	 & \raisebox{-2.2pt}{\!\includegraphics[scale=0.18]{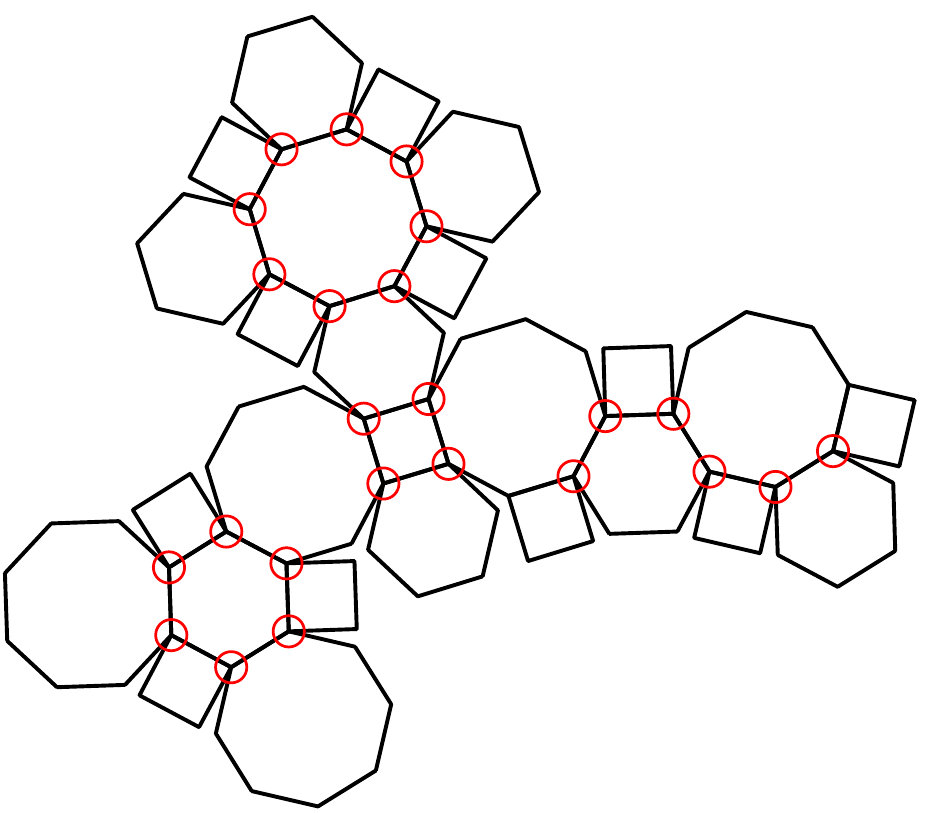}}
\\ \hline &&&&&& \\[-9.8pt]
\makecell{Truncated\\ Icosahedron \\ (soccer ball)}           	&	$60$& $32$ & $90$ & $30$ & $4\ 114$ & \raisebox{-2pt}{\!\includegraphics[scale=0.17,angle=90]{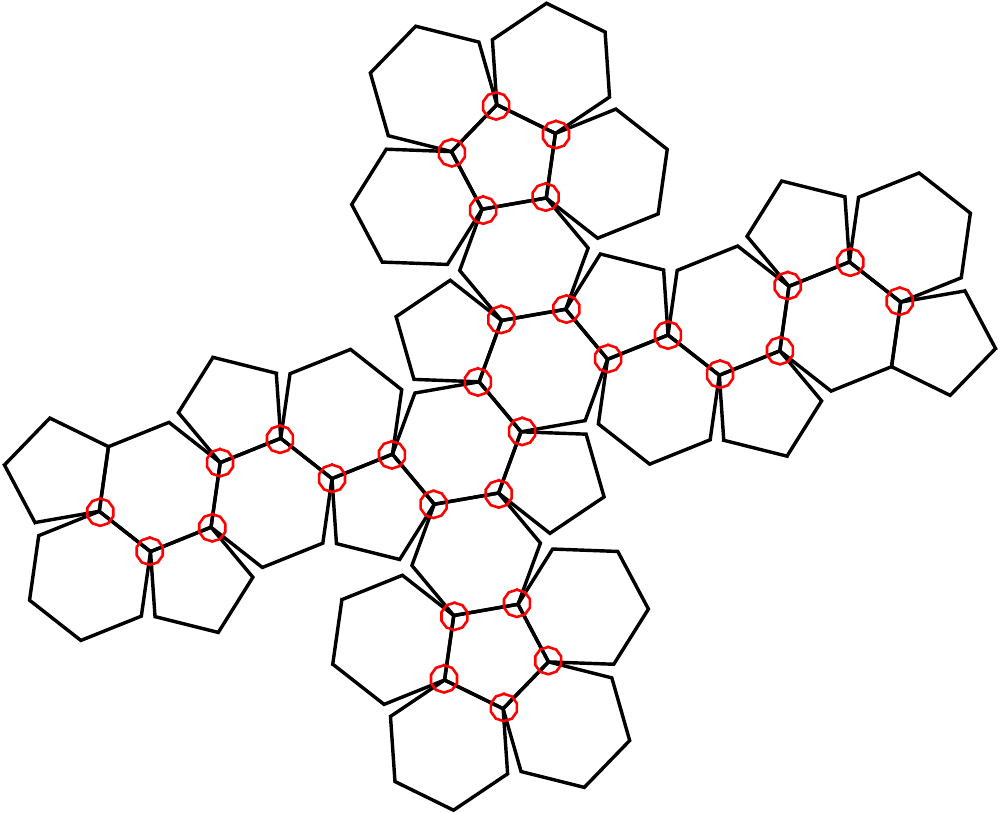}}
\\ \hline
\makecell{Truncated\\ Dodecahedron}         	&	$60$& $32$ & $90$ & $22$ & $3\ 719\ 677\ 167$ & --
\\ \hline
%Small Rhombicosidodecahedron 
\makecell{Rhombicosi-\\dodecahedron} &  	$60$& $62$ &$120$ & $37$ & $77\ 952$ &  --
\\ \hline
\makecell{Snub\\ Dodecahedron}                 &  	$60$& $92$ &$150$ & $39$ & $13\ 436\ 928$  & --
\\ \hline
\makecell{Triakis\\ Icosahedron}             	&	$32$& $60$ & $90$ & $26$ &  $664\ 128$	 & --
\\ \hline
\makecell{Pentakis\\ Dodecahedron}         	&	$32$& $60$ & $90$ & $22$ &  $845\ 280$	 & --
\\ \hline
\end{tabular}
\caption{Optimal nets. The number of leaves, $L$, in the MLSTs, and the number of distinct optimal nets (unlabeled MLSTs), $N_\text{opt nets}$, were obtained
for each polyhedron with our algorithm.
 The numbers of vertices $V$, faces $F$, and edges $E$ are also shown, as well as the optimal net with the smallest radius of gyration (only for cases with $N_\text{opt nets}<10\, 000$). The red circles indicate the vertex connections.
}
\label{t2}
\end{table}

\section{S3.\ \ \ Shells with holes}
\label{sec_holes}
We consider now the problem of finding optimal nets for shells that contain
holes, i.e., shells consisting of all the faces of a polyhedron except for
one.  The graph of the vertices and edges of the shell is the same as the
polyhedral graph of the complete polyhedron.  The difference is that all the
edges adjacent to the missing face will be in every cut, effectively detaching
that face from the rest of the net, as intended. (An edge adjacent to a face
is an edge between two vertices of that face.) For this reason, a vertex
adjacent to the missing face cannot be a vertex connection in the net,
since it always has two edges included in the cut.

The subgraph of the cut edges in the presence of a hole is not a pure tree,
because it contains a single loop formed by the edges adjacent to the hole. For
the shell to unfold into a 2D net, the cut subgraph must reach all vertices
(spanning) and be connected.  Also, for the shell's faces to remain connected
in a single component, the cut subgraph cannot have any other loop apart from
one surrounding the hole.  Then, the cut subgraph consists of the loop adjacent
to the hole, and some loopless branches connected to it.

We use the following Algorithm~\ref{hole} to maximize the number of vertex
connections in nets of shells that contain holes.  Algorithm~\ref{hole} is a
simple adaptation of Algorithm~\ref{main} that calls the same recursive
function of Algorithm~\ref{recursive}, gradually increasing the allowed number
of non-leaf vertices $n_S$.  The recursive function remains unchanged in this
procedure.  In this version of the algorithm, in addition to the lists of
adjacencies, $A=\{A_i\}$, we supply the list of vertices adjacent to the hole,
$V_h$.  Instead of single root vertex, the search starts from a subgraph
already containing all the vertices and edges adjacent to the hole (these
vertices and edges must be present in all the cuts, optimal or not).  Then, in
the first call of the function \textsc{recursive} for each $n_S$, $V_T=V_h$ and
$E_T$ is the set edges adjacent to the hole.  The set $E_A$ is initialized with
the edges connected to the vertices of $V_h$ that are not in $E_T$, as in
Algorithm~\ref{main}.  For the sake of clarity, while in Algorithm~\ref{main}
we denote the set of edges of the optimal cuts by $MLSTs$, in
Algorithm~\ref{hole} we denote it by $Cuts$ because in the presence of holes
the cuts contain a loop surrounding the hole, and they are no longer trees.
When Algorithm~\ref{hole} calls the function \textsc{recursive}, it passes an
empty set as the fifth input argument.  This set is the list of vertices that
are forced to be leaves, called $V_\text{excl}$ in \textsc{recursive}, which we
do not use in this algorithm.

\begin{algorithm}[H]
\begin{algorithmic}
\caption{\textit{Enumeration of all optimal cuts of shells with a hole.} 
This procedure initializes the necessary variables and calls the recursive function of Algorithm~\ref{recursive}.
The set $Cuts$ stores the spanning subgraphs that include all edges adjacent to the hole and maximize the number of leaves.
}
\label{hole}

\Procedure{list\_cuts\_with\_hole}{$A,V_h$}
\\
\textbf{Input:}\\
Lists of neighbors, $A_i$, of each vertex $i$ ($A=\{A_i\}$).\\
Vertices of the hole $V_h$.\\

	\State $Cuts=\emptyset$
	\State $n_S=|V_h|$
	\While{$Cuts=\emptyset$}
	
		\State $E_T={e_{ij}}: i,j \in V_h$
		\State $E_A={e_{ij}: j \in A_i \textbf{ and } i \in V_h  \textbf{ and }  j \notin V_h }$
		\State ${Cuts}' = \Call{recursive} {A,V_h, E_T, E_A, \emptyset, n_S}$
		\State $Cuts=\text{append}(Cuts,{Cuts}')$	

		\State $n_S=n_S+1$
	\EndWhile

\EndProcedure
\end{algorithmic}
\end{algorithm}

\section{S4.\ \ \ Estimations}

\subsection{A.\ \ \ Maximum number of leaves} \label{ss3}
The aim of this section is to estimate the number of leaves $L$ in a MLST in
terms of the simplest possible polyhedral parameters, preferably as a function
of the number of vertices $V$, or of edges $E$. The exact number $L$ depends
on the details of the graph, and its determination requires to actually solve
the maximum leaf spanning tree problem. However, we can obtain a simple
estimate for $L$ by considering a local optimization algorithm for finding
approximated maximum leaf spanning trees. 

This algorithm iteratively grows a tree by progressively attaching to it
vertices of the graph, until the tree becomes spanning (i.e. reaches all
vertices in the polyhedral graph), in the following way:

\indent (i) To seed the iterative process, connect the highest degree vertex to
all its neighbors.

\indent (ii) If the current tree is a spanning tree the algorithm reaches its
end. Otherwise, from the vertices already in the tree, select the one with the
highest number of neighbors still not in the tree, connect it to those
neighbors, and repeat step (ii).

In this algorithm, except for the highest degree vertex of step (i), every
vertex added to the tree starts as a leaf attached to a non-leaf vertex. The
neighbors of non-leaf vertices of the intermediate (non-spanning) trees are
guaranteed to be in the tree as well.  Furthermore, the number of
leaves in the final spanning tree is $L=V-n_\text{iter}-1$, where
$n_\text{iter}$ is the total number of iterations, and the $-1$ is due to the
initial step (i).  The number $n_\text{iter}$ depends on how many vertices are
added to the tree in each iteration.

As we show in the following, the number of vertices added to the tree at each
iteration turns out to be close to $2$, regardless of the details of the
polyhedral (shell) graph.  If the polyhedron has no triangular faces, a vertex
of degree $k$ selected in step (ii) contributes at most with $k-1$ new leaves,
because each vertex in the tree has at least one neighbor also in the tree.
However, if the faces are triangular, any two vertices connected to each other
share two common neighbors.  In this case, the selected vertex has at least $3$
neighbors already in the tree, namely, the parent non-leaf vertex plus two
vertices that are common neighbors with the parent vertex, therefore it
contributes at most with $k-3$ new leaves.

In convex polyhedra with regular faces (equilateral triangles, squares, regular
pentagons, etc), the sum of the internal angles attached to a vertex must be
smaller than $2\pi$.  This strongly constrains the types of faces that can be
attached to a vertex of degree $k$, and in particular the number of triangles.
Notice that $k$ cannot be smaller than $3$, and that $k=6$ is not feasible even
with just triangles (in that case the sum of angles is equal to $2\pi$).  Then,
let us consider the number of new vertices added to the tree when the vertex
selected in step (ii) has degree $k=3$, $4$, and $5$.  A vertex with $k=5$ is
mainly surrounded by triangles (at least $4$ out of $5$ faces must be
triangles), so, when selected in step (ii), most of the times it can contribute
with $k-3=2$ new leaves to the tree.  A vertex with $k=4$ must have between $1$
and $4$ triangles among its faces, and it can contribute with $1$ to $3$ new
leaves depending on the particular configuration of the faces.  Finally, a
vertex with $k=3$ can have any number of triangles between $0$ and $3$.  On the
one hand, if it has $3$ triangles, this vertex is never selected in step (ii)
because it has $0=k-3$ neighbors still not in the tree.  On the other hand, if
the vertex has no triangular faces attached, it can contribute with $k-1=2$ new
leaves.

Due to these geometrical constraints, the average number of new vertices added
to the tree at each iteration is essentially independent of the size and local
details of the polyhedral graph, and is close to $2$.  Furthermore, we expect
these arguments to qualitatively hold also for irregular convex polyhedra.
Even when the internal angles of a face are different from each other, their
sum is the same, and so, on average, the internal angles are the same as for a
regular face.

\begin{figure}[h]
\includegraphics[scale=0.4]{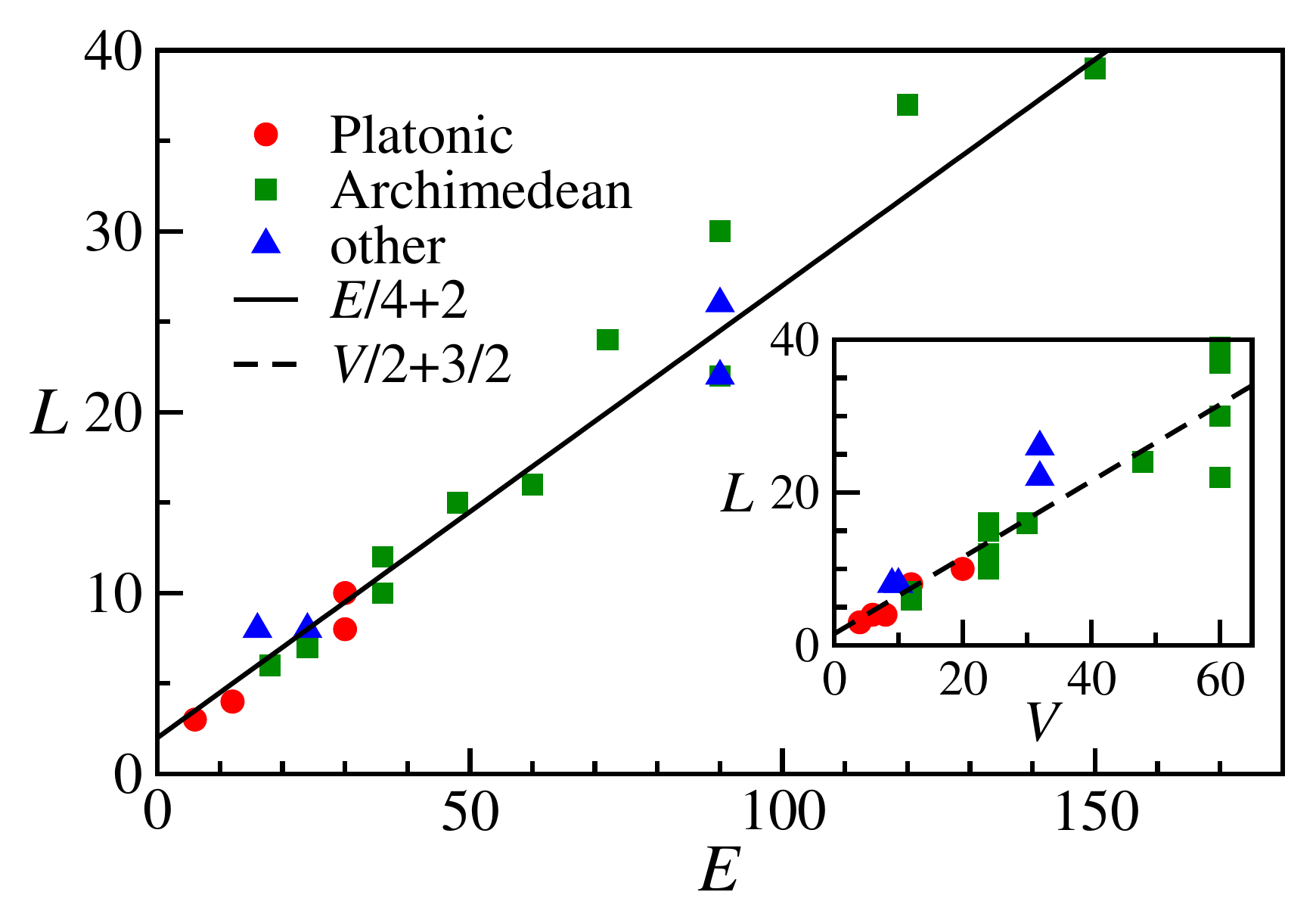}
\caption{
Number of leaf vertices in a maximum leaf spanning tree, $L$, vs.the number of edges $E$ of the polyhedra in Table~\ref{t2}. 
Different symbols represent different sets of polyhedra. 
The solid line is the estimation provided by Eq~(\ref{e30}). 
The inset shows $L$ vs. the number of vertices $V$, the dashed line is the estimation provided by Eq~(\ref{e20}).
}
\label{f2}
\end{figure}

Assuming that each iteration adds approximately $2$ new vertices on average to
the growing tree, and that the tree becomes spanning after $n_\text{iter}$
iterations, we can write:

\begin{equation}
k_0+1+ 2 n_\text{iter}  \sim V,
\label{e10}
\end{equation}
where $k_0$ is the highest degree in the polyhedral graph.
For simplicity, let us use $k_0=4$.
Replacing $n_\text{iter}=V-L-1$ in Eq.~(\ref{e10}) we get
\begin{equation}
L \sim (V+3)/2.
\label{e20}
\end{equation}
This formula fits well with the general trend observed for $L$ vs. $V$, as shown in the inset of Fig.~\ref{f2}.

Interestingly, the main panel of Fig.~\ref{f2} shows that the dispersion of the
points is significantly smaller if we plot $L$ vs $E$.  Let us recall Euler's
polyhedron formula $V+F=E+2$, where $V$, $F$, and $E$ are the numbers of
vertices, faces, and edges of the polyhedron, respectively. For a given $E$
there are different polyhedra with different combinations of $V$ and $F$ such
that $V+F=E+2$. On the one hand, if all faces are triangles we have $F=2E/3$
and in this case $V= E/3 + 2$.  On the other hand, if most faces of the
polyhedron have many edges (which implies wide internal angles) most vertices
are attached to only three faces, since the sum of the angles must be smaller
than $2\pi$ for convex polyhedra. Then, in this case $V\approx 2E/3$.  Taking
this into account, and because we consider a variety of polyhedra with
different types of faces, we use the middle point, 
\begin{equation}
V\sim E/2+1,
\label{e21}
\end{equation}
as an estimation of the number of vertices $V$ of a polyhedron with $E$ edges. 
We replace $E/2+1$ for $V$ in Eq.~(\ref{e20}), and finally obtain
\begin{equation}
L \sim E/4+2.
\label{e30}
\end{equation}
This formula, plotted in the main panel of Fig.~\ref{f2} as a solid line, shows a remarkable agreement with 
our results.

%%%%
%%%%
\subsection{B.\ \ \ Ratio $N_\text{MLST}/N_\text{ST}$}
%%%%
%%%%

We now estimate the ratio $N_\text{MLST}/N_\text{ST}$, between the
the numbers of maximum leaf spanning trees, $N_\text{MLST}$, and of spanning
trees, $N_\text{ST}$, in terms of the number of edges, $E$, of a labeled
polyhedral graph.  These numbers grow very quickly with the size of the graph
and are not uniquely determined by $E$ -- they depend on the details of the
graph.  To obtain a simple estimate, we calculate upper bounds for
$N_\text{ST}$ and $N_\text{MLST}$, and use their ratio as an estimator. 

Each spanning tree has $V-1$ edges out of a total of $E$ edges.  Therefore, the
upper bound of the number of spanning trees $N_\text{ST}$ is the number of
possible combinations of $V-1$ edges, given by the binomial coefficient
$\binom{E}{V-1}$.  Similarly, each maximum leaf spanning tree has $L$ leaves of
a total of $V$ vertices, and the upper bound for $N_\text{MLST}$ is the number
of combinations of $L$ vertices, $\binom{V}{L}$.

\begin{figure}[h]
\includegraphics[scale=0.4]{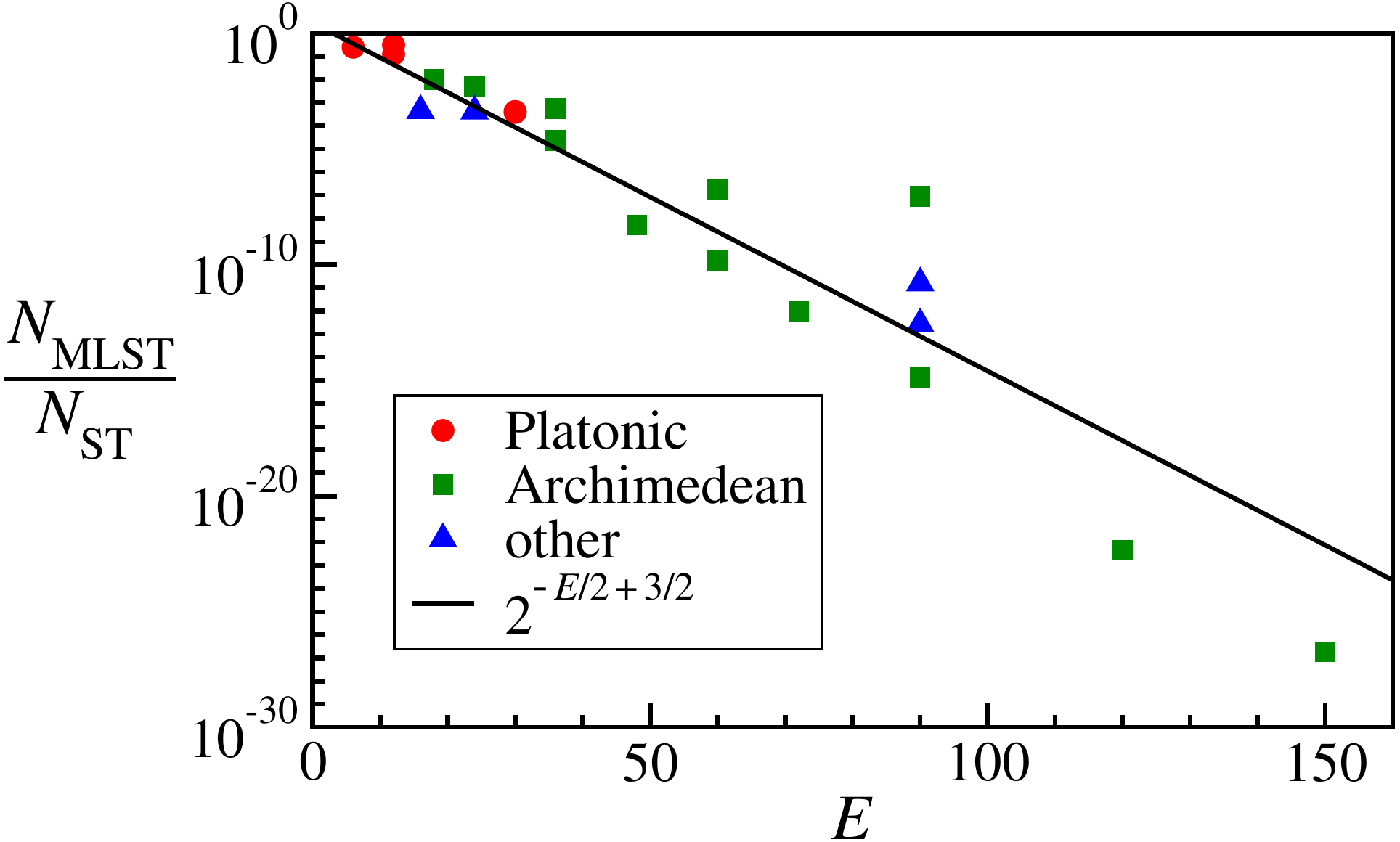}
\caption{Ratio between the total number of spanning trees, $N_\text{ST}$, and the number of maximum leaf spanning trees, $N_\text{MLST}$ vs. the number of edges $E$ of the polyhedra in Table~\ref{t2}. Different symbols represent different sets of polyhedra.
The ratio $N_\text{MLST}/N_\text{ST}$ decays exponentially with the polyhedron size, as predicted by Eq.~(\ref{e40}).
}
\label{f3}
\end{figure}

To obtain an estimation for $N_\text{MLST}/N_\text{ST}$, we take the ratio of
the upper bounds and replace the approximations for $L$ and $V$ of
Eqs.~(\ref{e20}) and~(\ref{e21}), respectively:
\begin{equation}
N_\text{MLST}/N_\text{ST}  \sim
\binom{E/2+1}{E/4+2} /  \binom{E}{E/2} 
  \sim 2^{-E/2+3/2} 
\label{e40}
\end{equation}
where we used Stirling's approximation $n! \sim \sqrt{2\pi n} (n/e)^n$.
Figure~\ref{f3} clearly shows that the ratio $N_\text{MLST}/N_\text{ST}$ has an
exponential-like decay with the number of edges of the polyhedron. The simple
form of Eq.~(\ref{e40}) fits well with this decay.

%%%%%%%%
%%%%%%%%
%%%%%%%%
\section{S5.\ \ \ Non-isomorphic cuts}  %############ ISOMORPHIC ############
%%%%%%%%
%%%%%%%%
%%%%%%%%
\label{auto}
To determine the optimal nets of a polyhedron, which have no labels, we need to
find the set of distinct unlabeled MLSTs. However, the algorithm of
Section~\ref{MLST} distinguishes each vertex by its label, and does not consider
symmetries that may exist in the polyhedral graph (automorphisms).  An
automorphism of a labeled graph is a relabeling that results in the same graph.
If the polyhedral graph contains automorphisms, then the set of MLSTs may
contain isomorphisms, i.e., multiple `copies' (differently labeled) of the same
unlabeled MLST~\cite{buekenhout1998number}.  These isomorphic cuts correspond
to nets that are indistinguishable from each other and have the same radius of
gyration, which is our second criterion of optimization. Therefore, we need
only one member of each set of isomorphic MLSTs in the list of optimal cuts.
To determine if two MLSTs are isomorphic we employ an adjacency matrix
approach. (Note that this approach is valid for any isomorphic subgraphs, not
just MLSTs.)

The adjacency matrix, $A$, is a convenient representation of labeled graphs.
Each element $A_{ij}$, is $1$ if the vertices with labels $i$ and $j$ are
connected by an edge, otherwise $A_{ij}$ is $0$.  To switch the labels of any
pair of vertices, say $i$ and $j$, we simply switch the rows $i$ and $j$ and
the columns $i$ and $j$ in matrix $A$.  Any permutation of $V$ labels can be
mapped into any other by a series of at most $V-1$ switches between pairs of
vertices ($V$ is the number of vertices).  We find the complete set of
automorphisms of the polyhedral graph by comparing the relabeled adjacency
matrix $A'$ with the original matrix $A$: when $A'=A$ the relabeling is
automorphic.

Similarly to the polyhedral graph, we can represent a MLST by an adjacency
matrix $B$ where each element $B_{ij}$ is $1$ if vertices $i$ and $j$ are
connected by an edge in the MLST and $0$ otherwise.  Two MLSTs, or cuts, are
isomorphic if, and only if, there is an automorphism of the polyhedral graph
that maps one MLST into the other.  That is, if two cuts, with adjacency
matrices $B_1$ and $B_2$, unfold into the same net, then there are automorphic
relabelings of $A$ that map $B_1$ into $B_2$ and vice-versa.  We apply each of
the previously obtained automorphisms of $A$ to one of the matrices, say $B_2$,
and compare the relabeled matrix $B_2'$ with $B_1$.  If one of those relabeling
gives $B_2'=B_1$, then the two cuts are isomorphic, and we may discard one of
them.  We systematically compare each of the MLSTs remaining in the list with
all the others to eliminate any isomorphisms, and obtain the full set of
$N_\text{opt net}$ distinct optimal nets of the polyhedron.

%%%%%%%%
%%%%%%%%
%%%%%%%%
\section{S6.\ \ \ Number of labeled spanning trees}
%%%%%%%%
%%%%%%%%
%%%%%%%%

Kirchhoff's matrix-tree theorem allows us to calculate the exact number of
spanning trees of a labeled graph, $N_\text{ST}$, in terms of the spectrum of
the Laplacian matrix.  The Laplacian matrix of a graph is defined as
$\mathcal{L}=D-A$, where $D$ is the degree matrix (i.e., the diagonal matrix
with each entry $d_{ii}$ equal to the degree of vertex $i$), and $A$ is the
adjacency matrix.  For a graph of $V$ vertices the matrix $\mathcal{L}$ has $V$
eigenvalues $\lambda_i\geq0$, the smallest of which is $\lambda_V=0$.  The
matrix-tree theorem states that the total number of spanning trees,
$N_\text{ST}$, is given by the product of eigenvalues of its Laplacian matrix
\begin{equation}
	N_\text{ST}=\frac{1}{V} \prod_{i=1}^{V-1}\lambda_i,
	\label{e6}
\end{equation}
where $\lambda_V$ is excluded from the product, and $\lambda_{i}>0$ for
connected graphs and $i<V$\cite{buekenhout1998number,cvetkovic1998spectra}.  The values
of $N_\text{ST}$ used to plot the points in Fig.~\ref{f3} (Fig.~3 of the main
text) were obtained with Eq.~(\ref{e6}) (while the values of $N_\text{MLST}$
were obtained by the algorithm described in Section~\ref{MLST}). 

\begin{table}[H]
\begin{tabular}{c | c | c | c | c | c }
  & $V$ & $E$ & $N_\text{nets}$ & {$N_\text{ST}/N_\text{aut}$} & $\Delta$  \\[0pt]
     \hline
  Tetrahedron & $4$ & $6$ & $2$ & $2/3$ & $0.66$ \\
  Cube & $8$ & $12$ & $11$ & $8$ & $0.27$ \\
  Octahedron & $6$ & $12$ & $11$ & $8$ & $0.27$ \\
  Icosahedron & $12$ & $30$ & $43\,380$ & $43\,200$ & $0.0041$\\
  Dodecahedron & $20$ & $30$ & $43\,380$ & $43\,200$ & $0.0041$\\
  $5$-cell  & $5$ & $10$ & $3$ & $25/24$ & $0.65$ \\
  $8$-cell & $16$ & $32$ & $261$ & $216$ & $0.17$ \\
  $16$-cell & $8$ & $24$ & $110\,912$ & $110\,592$ & $0.0029$ \\
  $24$-cell  & $24$ & $96$ & $1.79{\times}10^{16}$ & $1.79{\times}10^{16}$  & $1.9{\times}10^{-11}$ \\
  $120$-cell & $600$ & $1200$ & $2.76{\times}10^{119}$ & $2.76{\times}10^{119}$ & $3.9{\times}10^{-61}$\\
  $600$-cell  & $120$ & $720$ & ${1.20{\times}10^{307}}^*$ & $1.20{\times}10^{307}$ & $0$
\end{tabular}
\caption{Comparison of the exact number of nets $N_\text{nets}$ \cite{buekenhout1998number} and the estimation $N_\text{ST}/N_\text{aut}$ for all regular convex polytopes in $3$D (top 5 rows) and $4$D (bottom 6 rows). 
The right-hand side column of the table is the relative difference $\Delta=\left( N_\text{nets} - N_\text{ST}/N_\text{aut} \right) / N_\text{nets}$. 
The numbers of vertices, and edges are denoted $V$ and $E$, respectively. 
%The relative difference $\Delta$ approaches $0$ very quickly as the size of the polytope increases. 
We include the $4$D polytopes in this table to show that $\Delta$ approaches $0$ very quickly as the size ($V$ and $E$) of the polytope increases. 
Note that everywhere else in this paper we consider only $3$D shells and their $2$D nets.
\\$^*$The number of nets of the $600$-cell shown in Ref.~\cite{buekenhout1998number} is wrong, due to a mistake in the calculation of the graph's spectrum.
 Correcting the mistake gives  exactly \mbox{$2^{182}{\cdot} 3^{102}{\cdot} 5^{20}{\cdot} 7^{36}{\cdot} 11^{48}{\cdot} 23^{48}{\cdot} 29^{30}$} nets for the $600$-cell. 
 In this particular case, the lower-bound and the actual number of nets coincide exactly.
}
\label{t1}
\end{table}
%
%
%

%%%%%%%%
%%%%%%%%
%%%%%%%%
\section{S7.\ \ \ Number of non-isomorphic cuts}
%%%%%%%%
%%%%%%%%
%%%%%%%%
The number of non-isomorphic cuts (i.e. nets), $N_\text{nets}$, of a polyhedron
with no automorphisms is equal to $N_\text{ST}$.  When the polyhedral graph has
automorphisms, the number of distinct nets is actually smaller than
$N_\text{ST}$, see Section~\ref{auto}.  In that case, the exact number of nets,
$N_\text{nets}$, can be obtained using the approach of
Ref.~\cite{buekenhout1998number}, which involves a detailed, case by case,
analysis for each polyhedron.  In Ref.~\cite{buekenhout1998number} the
$N_\text{nets}$ are obtained only for the five platonic solids.  Nevertheless,
we can estimate $N_\text{nets}$ for other polyhedron with automorphisms (e.g.,
Archimedean solids) with high precision by taking the ratio of $N_\text{ST}$ to
the number of automorphisms of the graph, $N_\text{aut}$.  The ratio
$N_\text{ST}/N_\text{aut}$ is in fact a lower-bound of $N_\text{nets}$; This
ratio assumes that each unlabeled spanning tree contributes with $N_\text{aut}$
differently labeled `copies' to the set of labeled spanning trees, however,
some spanning trees have a smaller number of isomorphic `copies', due to the
existence of symmetries in their branches.  Since $N_\text{aut}$ can be found
with linear time algorithms \cite{hopcroft1974linear}, the calculation of
$N_\text{ST}/N_\text{aut}$ is straightforward.

\begin{figure}[ht]
\ \includegraphics[scale=0.4]{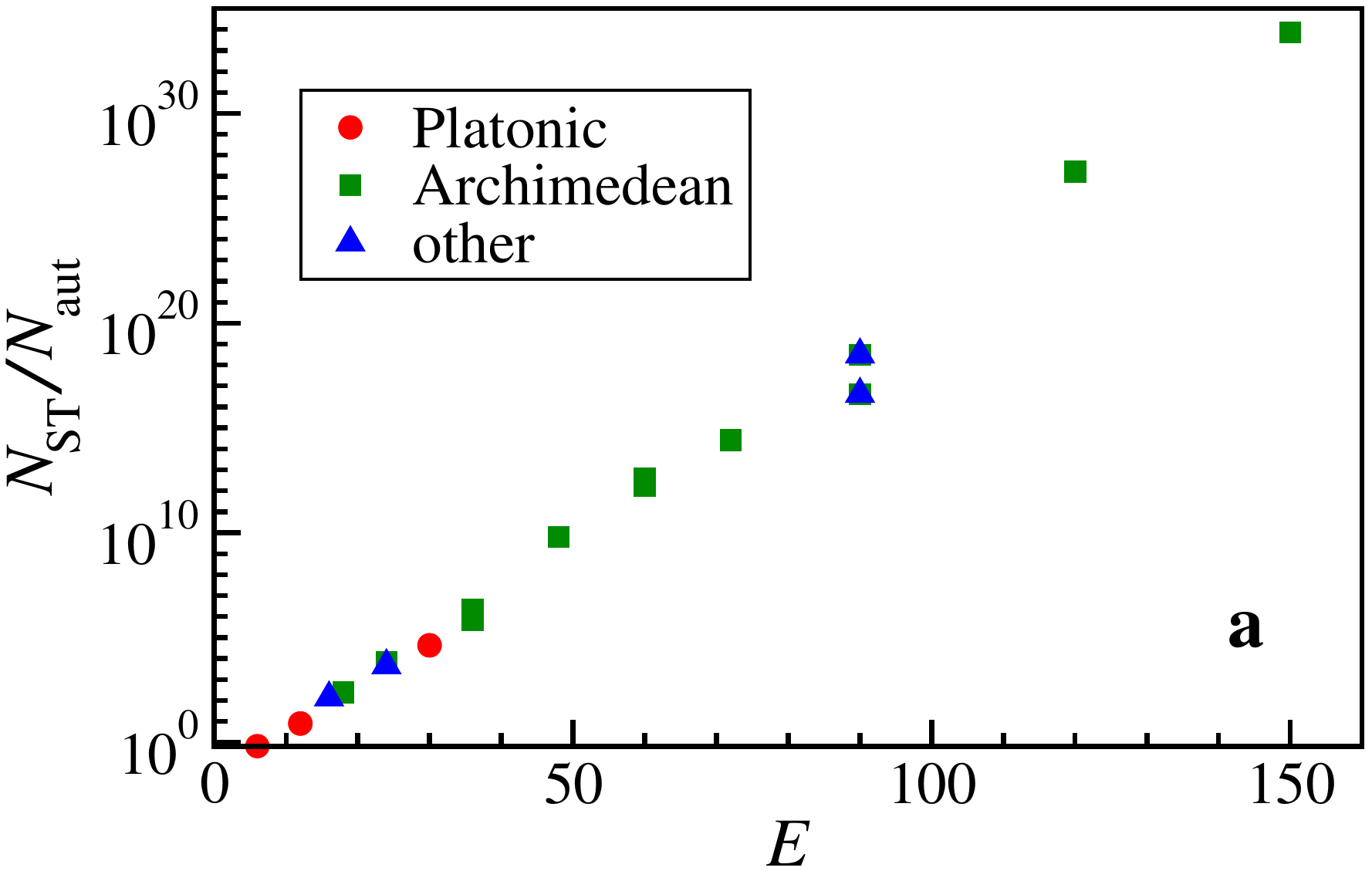}
\\
\hspace{-3.7pt} \includegraphics[scale=0.4]{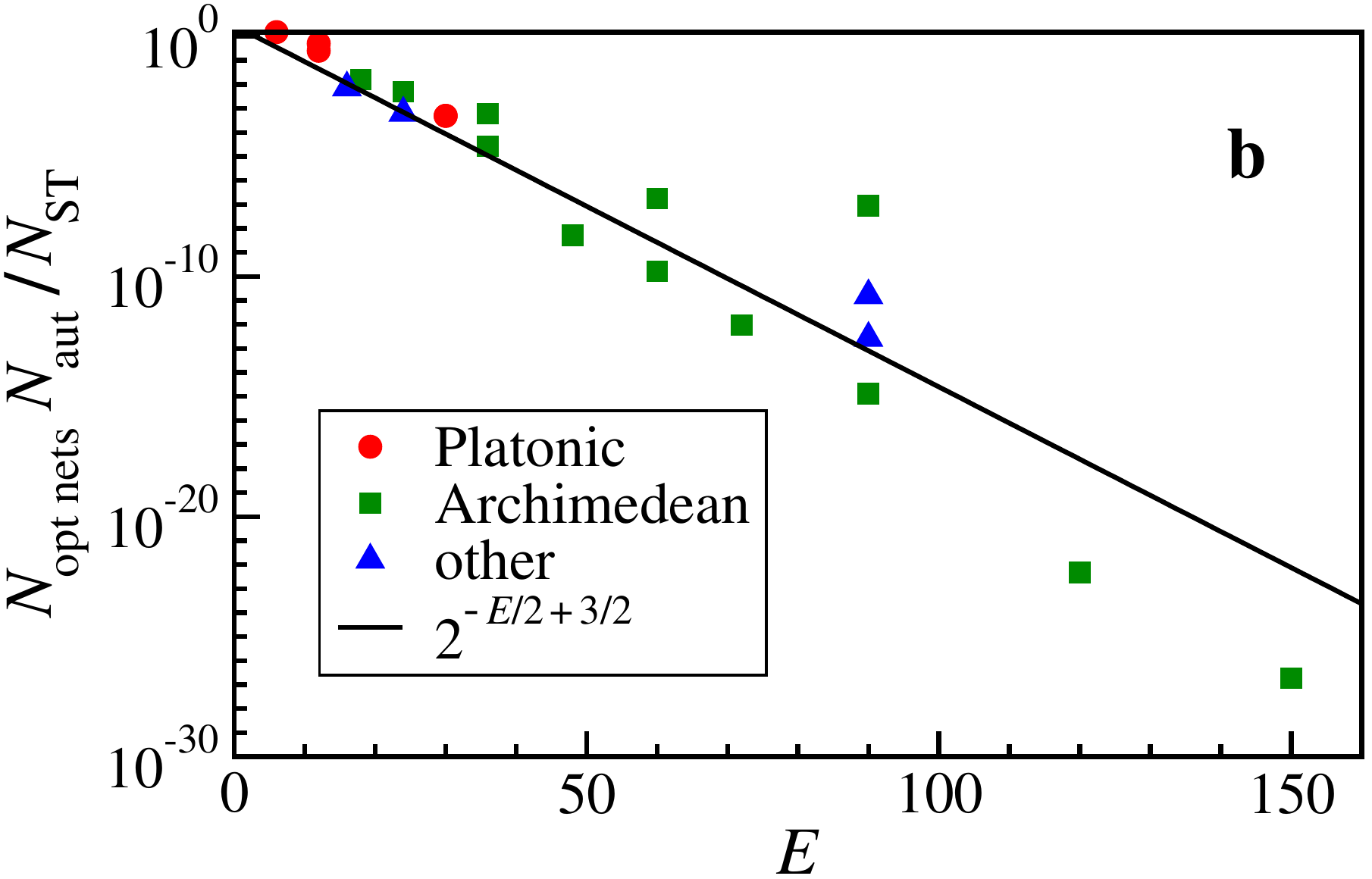}
\caption{
Precise estimations of (a) the number of distinct nets $N_\text{nets}\approx N_\text{ST}/N_\text{aut}$, and of (b) the fraction of optimal nets  $ N_\text{opt nets} /N_\text{nets} \approx N_\text{opt nets} N_\text{aut}/N_\text{ST}$ vs. the number of edges $E$ of the $3$D polyhedra in Table~\ref{t2}. 
The number $N_\text{nets}$ grows exponentially with $E$, and shows a remarkably low level of dispersion. 
The solid line in the plot of panel (b) is Eq.~(\ref{e40}). 
The fraction of unlabeled optimal cuts $N_\text{opt nets} /N_\text{nets}$ is essentially equal to the fraction of labeled optimal cuts, $N_\text{MLST}/N_\text{ST}$, see Fig.~\ref{f3}.
}
\label{f1}
\end{figure}

Table~\ref{t1} clearly shows that this lower-bound actually approaches
$N_\text{nets}$ very quickly for large graphs.  This happens because the
fraction of spanning trees that have some symmetry in their structure
quickly approaches $0$ as the size of the graph increases.  Figure~\ref{f1}(a)
demonstrates that the number of distinct nets, $N_\text{nets} \approx
N_\text{ST}/N_\text{aut}$, grows exponentially with the size of the polyhedral
graph.

%%%%%%
%%%%%%
%%%%%%
%%%%%%

The probability that a randomly sampled labeled spanning tree is a MLST is
equal to the ratio $N_\text{MLST}/N_\text{ST}$, which is shown is Fig.~3 of the
main text, and reproduced in this Supplemental Material as Fig.~\ref{f3}.  If
we were to generate nets with a method that avoids the sampling of isomorphic
cuts completely, thus reducing the search space, the probability that a
randomly sampled net has the maximum number of vertex connections,
$N_\text{opt nets}/N_\text{nets}$, would still be essentially the same as
$N_\text{MLST}/N_\text{ST}$. 

Figure~\ref{f1}(b) shows the probability that a net sampled at random from the
set of non-isomorphic nets has the maximum number of vertex connections,
i.e., $N_\text{opt nets}/N_\text{nets} \approx N_\text{opt nets}\,
N_\text{aut}/N_\text{ST}$.  We observe no significant differences between
Figs.~\ref{f3}~and~\ref{f1}(b), which means that searches over unlabeled and
labeled configurations have similar performances when only a small fraction of
the configurations is sampled.

%\bibliography{Supp_Mat.bib}

%

\section{S8.\ \ \ Source code for closed shells}
In this sections we print the source code of a Mathematica implementation of the algorithm presented in Sec.~S2.
This is the code that can be found in the file \verb|Procedure1_for_closed_shells.nb| available at URL \url{https://journals.aps.org/prl/supplemental/10.1103/PhysRevLett.120.188001}.
The inputs supplied to this Mathematica routine are the adjacency matrix of the shell graph, the lists of vertices adjacent to each face, and the spatial coordinates of each vertex in the polyhedron. 
During its execution the script outputs informations about the progress of the algorithm, and returns a list of all the nets with the maximum number of vertex connections.
This list is sorted by the radius of gyration. 
The script also produces a plot of the ranked radii of gyration for
 the optimal nets found by the algorithm.
The particular polyhedron
\linebreak
\newpage
\noindent
 treated by script printed here is the structure presented in Fig.~2c of the main paper, the Small Rhombicuboctahedron.

\includepdf[pages={1},pagecommand=\thispagestyle{headings}]{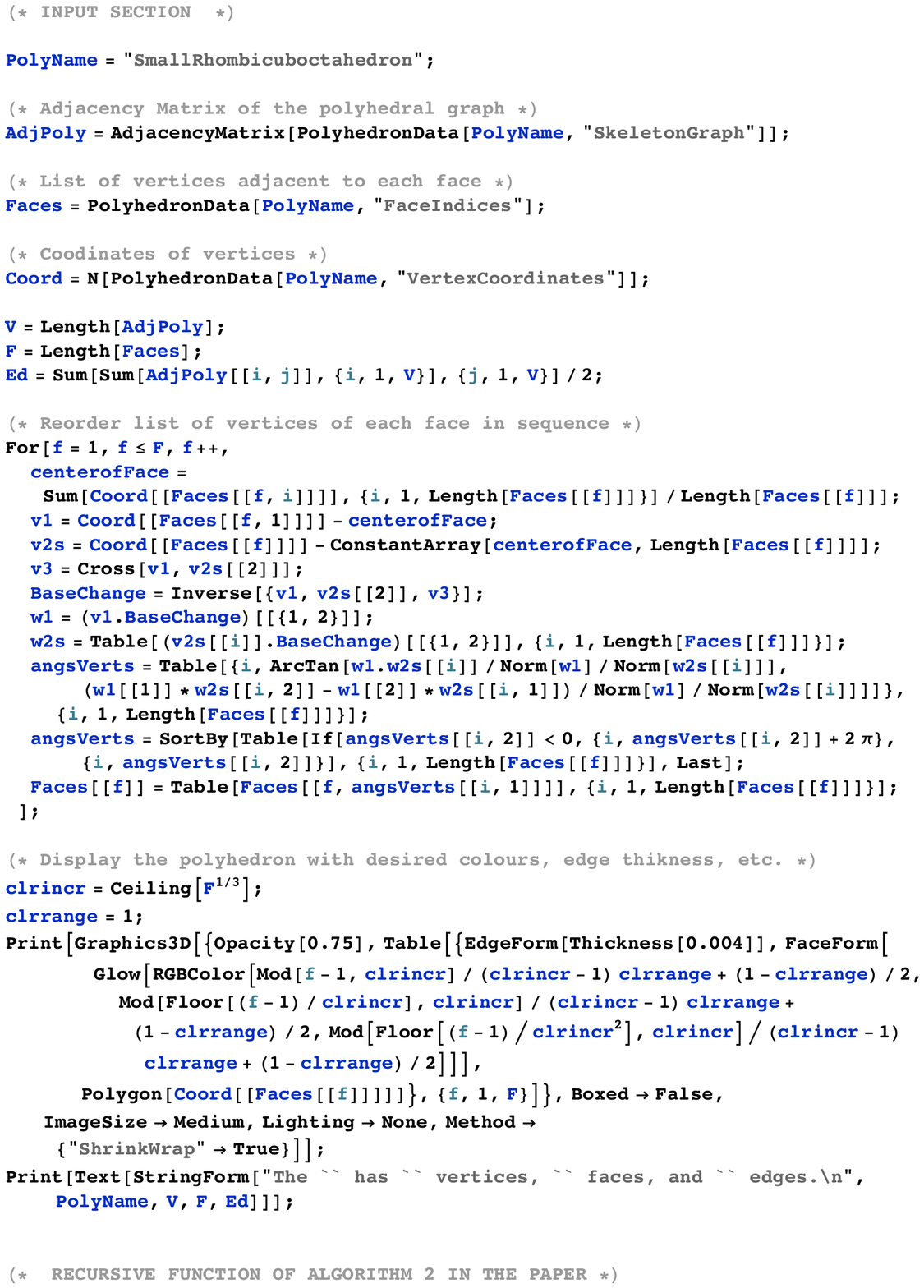}
\includepdf[pages={{},2},pagecommand=\thispagestyle{headings}]{Procedure1_for_closed_shells.pdf}
\includepdf[pages={{},3},pagecommand=\thispagestyle{headings}]{Procedure1_for_closed_shells.pdf}
\includepdf[pages={{},4},pagecommand=\thispagestyle{headings}]{Procedure1_for_closed_shells.pdf}
\includepdf[pages={{},5},pagecommand=\thispagestyle{headings}]{Procedure1_for_closed_shells.pdf}
\includepdf[pages={{},6},pagecommand=\thispagestyle{headings}]{Procedure1_for_closed_shells.pdf}
\includepdf[pages={{},7},pagecommand=\thispagestyle{headings}]{Procedure1_for_closed_shells.pdf}
\includepdf[pages={{},8},pagecommand=\thispagestyle{headings}]{Procedure1_for_closed_shells.pdf}
\includepdf[pages={{},9},pagecommand=\thispagestyle{headings}]{Procedure1_for_closed_shells.pdf}
\includepdf[pages={{},10},pagecommand=\thispagestyle{headings}]{Procedure1_for_closed_shells.pdf}
\includepdf[pages={{},11},pagecommand=\thispagestyle{headings}]{Procedure1_for_closed_shells.pdf}
\includepdf[pages={{},12},pagecommand=\thispagestyle{headings}]{Procedure1_for_closed_shells.pdf}
\includepdf[pages={{},13},pagecommand=\thispagestyle{headings}]{Procedure1_for_closed_shells.pdf}
\includepdf[pages={14},pagecommand=\thispagestyle{headings}]{Procedure1_for_closed_shells.pdf}

\pagebreak

\section{S9.\ \ \ Source code for open shells}
 In this sections we print the source code of a Mathematica implementation of the algorithm presented in Sec.~S3.
This is the code that can be found in the file \verb|Procedure3_for_open_shells.nb| available at URL \url{https://journals.aps.org/prl/supplemental/10.1103/PhysRevLett.120.188001}.
This routine for shells with a hole needs the same inputs as the one for closed shells, and, additionally, the list of the vertices adjacent to the open
 face(s).
During its execution the script outputs informations about the 
\linebreak
\newpage
\section{ }
\noindent
progress of the algorithm, and returns a list of all the nets with the maximum number of vertex connections.
This list is sorted by the radius of gyration. 
The script also produces a plot of the ranked radii of gyration for the optimal nets found by the algorithm.
The particular open shell treated by script printed here is the structure presented in Fig.~2e of the main paper, the Small Rhombicuboctahedron with the top nine faces removed.

\includepdf[pages={1},pagecommand=\thispagestyle{headings}]{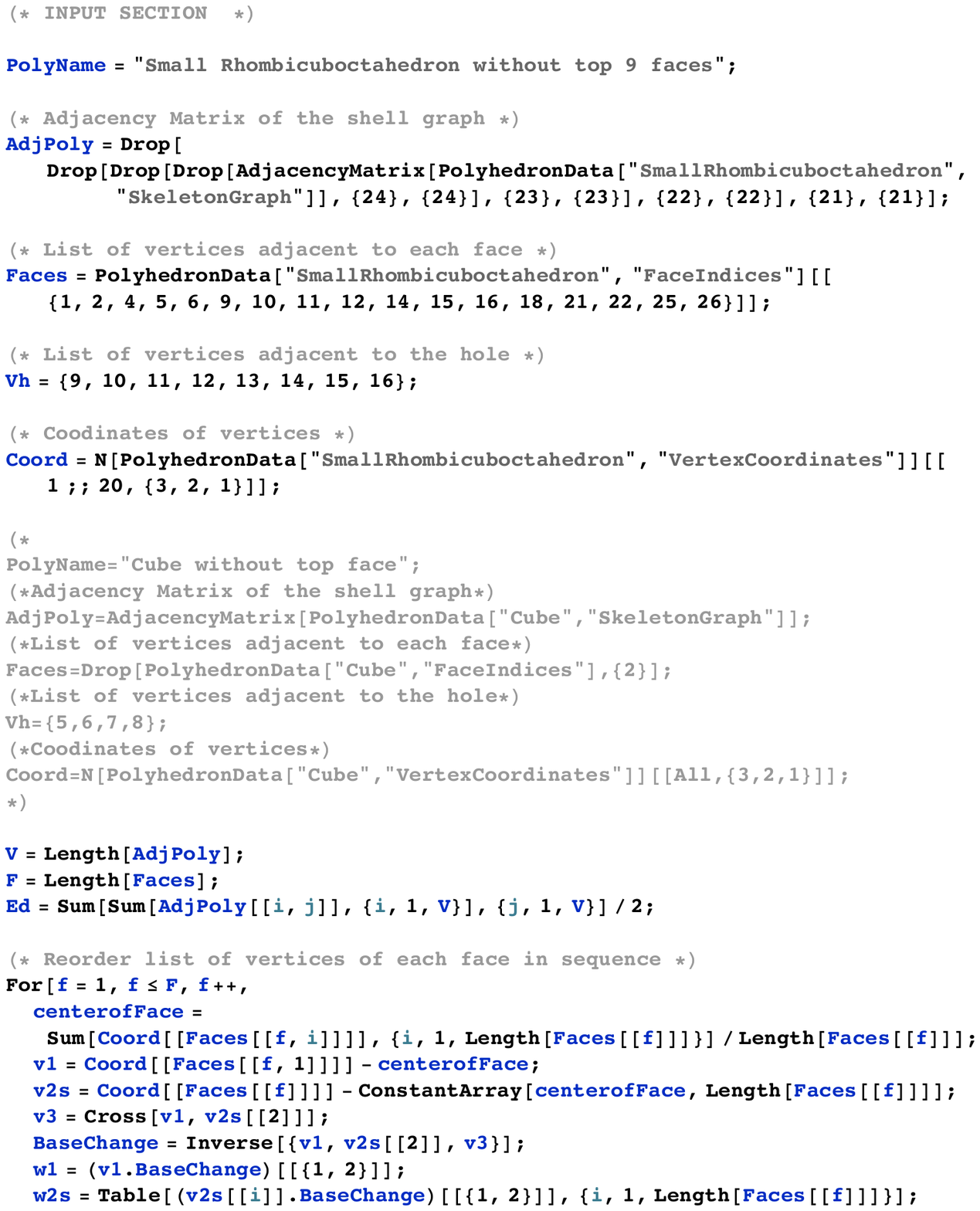}
\includepdf[pages={{},2},pagecommand=\thispagestyle{headings}]{Procedure3_for_open_shells.pdf}
\includepdf[pages={{},3},pagecommand=\thispagestyle{headings}]{Procedure3_for_open_shells.pdf}
\includepdf[pages={{},4},pagecommand=\thispagestyle{headings}]{Procedure3_for_open_shells.pdf}
\includepdf[pages={{},5},pagecommand=\thispagestyle{headings}]{Procedure3_for_open_shells.pdf}
\includepdf[pages={{},6},pagecommand=\thispagestyle{headings}]{Procedure3_for_open_shells.pdf}
\includepdf[pages={{},7},pagecommand=\thispagestyle{headings}]{Procedure3_for_open_shells.pdf}
\includepdf[pages={{},8},pagecommand=\thispagestyle{headings}]{Procedure3_for_open_shells.pdf}
\includepdf[pages={{},9},pagecommand=\thispagestyle{headings}]{Procedure3_for_open_shells.pdf}
\includepdf[pages={{},10},pagecommand=\thispagestyle{headings}]{Procedure3_for_open_shells.pdf}
\includepdf[pages={{},11},pagecommand=\thispagestyle{headings}]{Procedure3_for_open_shells.pdf}
\includepdf[pages={{},12},pagecommand=\thispagestyle{headings}]{Procedure3_for_open_shells.pdf}
\includepdf[pages={{},13},pagecommand=\thispagestyle{headings}]{Procedure3_for_open_shells.pdf}
\includepdf[pages={{},14},pagecommand=\thispagestyle{headings}]{Procedure3_for_open_shells.pdf}
\includepdf[pages={{},15},pagecommand=\thispagestyle{headings}]{Procedure3_for_open_shells.pdf}
\includepdf[pages={{},16},pagecommand=\thispagestyle{headings}]{Procedure3_for_open_shells.pdf}
\includepdf[pages={{},17},pagecommand=\thispagestyle{headings}]{Procedure3_for_open_shells.pdf}
\includepdf[pages={{},18},pagecommand=\thispagestyle{headings}]{Procedure3_for_open_shells.pdf}
\includepdf[pages={{},19},pagecommand=\thispagestyle{headings}]{Procedure3_for_open_shells.pdf}

\end{document}